\documentclass[11pt,a4paper]{article}

\usepackage[hidelinks]{hyperref}
\usepackage{amssymb, amsmath, graphicx, subcaption}
\usepackage{amsthm}
\usepackage{autobreak}
\usepackage[ruled]{algorithm2e}
\usepackage{multirow}
\usepackage{float}
\usepackage{fancyhdr}

\usepackage{geometry}
\fancypagestyle{firststyle}
{
	\fancyhf{}
	\fancyhead[RE,LO]{}
	\fancyfoot[RE,LO]{}
	\fancyfoot[LE,RO]{Page \thepage}
}

\begin{document}
	
	\title{Analytic transformation from osculating to mean elements under J2 perturbation}
	
	\author{David Arnas\thanks{Purdue University, West Lafayette, IN 47907, USA. Email: \textsc{darnas@purdue.edu}}}
	
	\date{}	
	
	\maketitle
	
	\thispagestyle{firststyle}
	
	\begin{abstract}
		This work presents an analytical perturbation method to study the dynamics of an orbiting object subject to the term $J_2$ from the gravitational potential of the main celestial body. This is done using a power series expansion in the perturbation constant $J_2$ on all the variables of the system, and a time regularization based on the argument of latitude of the orbit. This enables the generation of analytic solutions without the need to control the perturbed frequency of the system. The resultant approach allows to approximate the dynamics of the system in osculating elements for orbits at any eccentricity, and to obtain the approximate analytical transformation from osculating to mean elements in these orbits. This includes near circular, elliptic, parabolic and hyperbolic orbits at any inclination. Several examples of application are presented to show the accuracy of the perturbation approach and their related transformations.
	\end{abstract}

\section{Introduction}

In astrodynamics, space flight mechanics, mission analysis and space situation awareness problems, it is of extreme importance to evaluate the orbits of spacecraft and any other orbiting object not only for a specific instant, but also for its long-term dynamics. In general, mean orbital elements provide a better insight about an orbit because they are easier to relate to our knowledge about unperturbed keplerian dynamics. However, the instantaneous state of an orbiting object only provides direct information about the osculating orbital elements and thus, it is usually required to perform an orbital propagation and average the results for a complete revolution in order to obtain the sought mean orbital elements. An alternative to that is to use analytical approximations to the problem.

Approximate analytical solutions for the $J_2$ problem have been extensively studied over the years to provide a better insight on both the short and long term dynamics of satellites subjected to this perturbation. This is primarily due to the fact that for satellites orbiting the Earth, the $J_2$ term of the gravitational potential is three orders of magnitude larger than any other term of the geopotential, and also much larger than the effect of other perturbations for satellites in Low Earth Orbit (LEO). For this reason, the $J_2$ problem is also known as the main satellite problem. However, even being the $J_2$ problem the simplest term to study in the geopotentital, it does not have an analytical solution~\cite{irigoyen1993non,celletti1995non}. Therefore, different perturbation techniques have appeared to approximate the dynamics under this perturbation. 

One of the first examples of this kind of approach was given by Brower~\cite{brouwer}, who proposed a first order solution based on the von Zeipel method. This solution presented several singularities for orbits with small eccentricities and inclinations, and orbits close to the critical inclination that were originated due to a perturbation method based on averaging techniques. In order to address them, Lyddane~\cite{lyddane1963small}, and Cohen and Lyddane~\cite{cohen1981radius} proposed solutions to remove the singularities related with small eccentricities and inclinations respectively, and afterwards, Coffey, Deprit and Miller~\cite{coffey1986critical} did the same to study orbits close to the critical inclination. Nearly at the same time that Brouwer's solution was published, Kozai~\cite{kozai1959motion} proposed an alternative first order solution based on a decomposition of the analytical solution into first-order secular, second-order secular, short-periodic, and long-periodic terms. Some years later, Kozai~\cite{kozai1962second} extended Brower's solution to second order, which significantly improved the accuracy of the approximate solution. Finally, Hori~\cite{hori}, based also on the von Ziepel method, provided a first order solution for the hyperbolic motion. 

Deprit~\cite{deprit1969}, on the other hand, introduced a new perturbation method based on Lie series transformations combined with a power series in the small parameter. These transformations define a set of recursive canonical mappings in the Hamiltonian of the system that allow to obtain the approximate solution to an arbitrary order. This led to the development of the Lie-Deprit methods, which have become one of the most used perturbation techniques in the literature~\cite{kamel1969expansion,kamel1970perturbation,deprit1970main,deprit1982delaunay,cid,abad2001short,lara2014delaunay,mahajan2018exact,lara2019new,abad2020integration,abad2021cid}. Based on this approach, for instance, the third order solution for the main satellite problem was obtained for the first time by Coffey and Deprit~\cite{3rdorder} for any elliptic orbit, using elimination of parallax~\cite{deprit1981elimination} as well as additional short and long period eliminations based on Lie transforms. 

The previous methodologies rely on performing, at some point, an averaging process either directly on the Hamiltonian, or on the system of differential equations. This significantly simplifies the resultant expressions obtained from the perturbation methods, but introduces some limitations especially when studying limit cases as parabolic orbits or when defining frozen orbits from an osculating perspective. Conversely, perturbation methods based on osculating orbital elements can address these limitations. Unfortunately, due to the size of their solutions, osculating perturbation methods are more challenging to extend to higher order approximate solutions, which has limit their use. Examples of this kind of approach include the application of the Poincar\'e-Lindstedt method~\cite{zonal} to the main satellite problem, or the use of operation theory~\cite{koopman,schur}. 

In this work, we follow this trend by studying the main satellite problem using osculating elements instead of averaging techniques. This is done using a perturbation method based on a power series expansion directly on the osculating elements. However, and compared with a Poincar\'e-Lindstedt method, there is no condition imposed on the perturbed frequency of the solution. Instead, the independent variable of the differential equation is transformed to reduce the effect of this simplification. This approach allows to obtain an accurate approximate solution of the system while maintaining a relatively low complexity of both the perturbation method and the resultant expressions. This result is used to define an approximate analytical transformation from osculating to mean elements for orbits at any eccentricity, which is the focus of this work. Additionally, and due to the osculating nature of the perturbation method and the solution, the proposed approach is able to define and study parabolic orbits under the $J_2$ perturbation, which is a problem specially difficult to study with other existing analytical approaches~\cite{lara2022torsion}.

This manuscript is organized as follows. First, a set of orbital elements is selected such that they do not present any singularity due to eccentricity. This allows to study both near circular orbits, and parabolic and hyperbolic orbits using the same set of orbital elements. Second, a variable transformation is proposed from this set of orbital elements to spherical coordinates. This is used to define the exact transformation of the differential equation for its use alongside this set of orbital elements. Third, the perturbation approach is presented. This contains a time regularization based on the osculating argument of latitude of the orbit and a power series expansion of all the variables involved in the problem. Based on this expansion, the approximate solution of the main satellite problem is obtained. Fourth, the transformation from osculating to mean elements is included for the different approximate orders of the solution. Finally, a set of examples are presented to analyze the expected accuracy and performance of the proposed solution both for short-term and long-term dynamics. This includes examples for near circular, highly elliptic, hyperbolic, and parabolic orbits. 

All the solutions provided in this document have been translated into Matlab code for their ease of use. The resultant scripts can be found in the following web page:
\href{https://engineering.purdue.edu/ART/research/research-code}{https://engineering.purdue.edu/ART/research/research-code}.


\section{The main satellite problem}

The resultant dynamics of an object subject to the oblatness of a celestial body (represented by the J2 term of its gravitational potential) can be represented using hamiltonian formulation in spherical coordinates. Let $r$ be the distance from the center of the celestial body to the orbiting object, let $\varphi$ be the latitude position of the orbiting object with respect to the celestial body's equator, and let $\lambda$ be the longitude position of that object with respect to the $x$ direction in an inertial frame of reference selected where the $x$ and $y$ axis are contained in the celestial body's equatorial plane. Then, the Hamiltonian of the system is:
\begin{equation}
	\mathcal{H} = \displaystyle\frac{1}{2}\left(p_r^2 + \frac{p_{\varphi}^2}{r^2} + \frac{p_{\lambda}^2}{r^2\cos^2(\varphi)}\right) - \frac{\mu}{r} + \frac{1}{2}\mu R^2J_2\frac{1}{r^3}\left(3\sin^2(\varphi) - 1\right),
\end{equation}
where:
\begin{eqnarray}
	p_r & = & \dot{r}; \nonumber \\
	p_{\varphi} & = & r^2\dot{\varphi}; \nonumber \\
	p_{\lambda} & = & r^2\cos^2(\varphi)\dot{\lambda};
\end{eqnarray}
$\mu$ is the celestial body's gravitational constant, and $R$ is the mean radius of the celestial body at its Equator. From this expression, it is possible to obtain the associated Hamilton's equations:
\begin{eqnarray}
	\displaystyle\frac{dr}{dt} & = & p_r; \nonumber \\
	\displaystyle\frac{dp_r}{dt} & = & -\frac{\mu}{r^2} + \frac{p_{\varphi}^2}{r^3}  + \frac{p_{\lambda}^2}{r^3\cos^2(\varphi)} + \frac{3}{2}\mu J_2 R^2\frac{1}{r^4}\left(3\sin^2(\varphi) - 1\right); \nonumber \\
	\displaystyle\frac{d\varphi}{dt} & = & \frac{p_{\varphi}}{r^2}; \nonumber \\
	\displaystyle\frac{dp_{\varphi}}{dt} & = & -\frac{p_{\lambda}^2}{r^2}\frac{\sin(\varphi)}{\cos^3(\varphi)} - 3\mu J_2 R^2\frac{1}{r^3}\sin(\varphi)\cos(\varphi); \nonumber \\
	\displaystyle\frac{d\lambda}{dt} & = & \frac{p_{\lambda}}{r^2\cos^2(\varphi)}; \nonumber \\
	\displaystyle\frac{dp_{\lambda}}{dt} & = & 0.
\end{eqnarray}
These equations are highly non-linear, making them specially challenging to use alongside analytic methodologies. For this reason, it is first necessary to transform the system into a differential equation with a linear unperturbed term, and a non-linear perturbation. This significantly helps in the generation of approximate analytical solutions.


\section{Variable transformation}

Keplerian elements are used in this work as a base to define the motion of orbits, namely, the semi-major axis of the orbit ($a$), the eccentricity ($e$), the argument of perigee ($\omega$), the inclination ($i$), the right ascension of the ascending node ($\Omega$), and the true anomaly ($\nu$). However, in order to avoid some of the singularities that this set of elements have and to simplify the resultant expressions, this work makes use of a modified set of orbital elements. 

In particular, the eccentricity and the argument of perigee are substituted by the components of the eccentricity vector in the orbital plane {$e_x, e_y$}:
\begin{eqnarray}
    e_x & = & e \cos(\omega), \nonumber \\
    e_y & = & e \sin(\omega).
\end{eqnarray}
This removes the problem of definition for near circular orbits and, in addition, it will simplify the first and second order solutions of the differential equation. Following the same reasoning, the true anomaly $\nu$ is substituted by the argument of latitude $\theta = \nu + \omega$. 

On the other hand, the semi-major axis is also substituted by a new variable $A$ which is defined as:
\begin{equation}\label{eq:A_def}
    A = \displaystyle\frac{R^2}{a^2(1-e^2)^2} = \frac{R^2 \mu^2}{p_{\theta}^4}.
\end{equation}
This variable change is motivated by three reasons. First, it avoids the singularity when dealing with parabolic orbits (note that $A$ is related with the inverse of the magnitude of the angular momentum $p_{\theta}$). Second, and as in the case of the eccentricity, this change of variable also reduces the length of the resultant solution from the perturbation method used in this work. And third, this provides a additional non-dimensional variable to the system, making all the set of orbital elements selected $\{A,e_x,e_y,i,\Omega,\theta \}$ non-dimensional.

Once this set of orbital elements $\{A,e_x,e_y,i,\Omega,\theta \}$ is defined, it is required to rewrite, in an exact form, the system of differential equations in terms of this modified set of variables. To that end, the following subsections cover these bijective transformations as well as the resultant system of differential equations.

\subsection{From spherical to orbital elements}

The first goal then is to find a transformation from spherical coordinates ($r$, $p_r$, $\varphi$, $p_{\varphi}$, $\lambda$, $p_{\lambda}$) to the chosen orbital elements ($A$, $e_x$, $e_y$, $i$, $\Omega$, $\theta$). We know that the magnitude of the angular momentum $p_\theta$ is equal to:
\begin{equation} \label{eq:angular_momentum}
    p_{\theta} = \displaystyle\sqrt{p_{\varphi}^2 + \frac{p_{\lambda}^2}{\cos^2(\varphi)}} = \sqrt{\mu a (1 - e^2)},
\end{equation}
which combined with the definition of A (Eq.~\eqref{eq:A_def}) leads to:
\begin{equation}
    A = \left(\displaystyle\frac{\mu R \cos^2(\varphi)}{p_{\varphi}^2\cos^2(\varphi) + p_{\lambda}^2}\right)^2.
\end{equation}

The semi-major axis of the orbit can be obtained trough the equation of energy for the unperturbed problem:
\begin{equation}
	\displaystyle\frac{1}{2}\left(p_r^2 + \frac{p_{\varphi}^2}{r^2} + \frac{p_{\lambda}^2}{r^2\cos^2(\varphi)}\right) - \frac{\mu}{r} = - \frac{\mu}{2a},
\end{equation}
where the inverse of the semi-major axis is:
\begin{equation} \label{eq:semi-major}
    \displaystyle\frac{1}{a} = \frac{2}{r} - \frac{p_r^2}{\mu} - \frac{p_{\theta}^2}{\mu r^2}.
\end{equation}
From the equation relating radial distance and true anomaly:
\begin{equation} \label{eq:radial}
    r = \displaystyle\frac{\frac{p_{\theta}^2}{\mu}}{1 + e\cos(\nu)},
\end{equation}
it is possible to obtain the following relation:
\begin{equation} \label{eq:ecos}
    e\cos(\nu) = e_x\cos(\theta) + e_y\sin(\theta) = \displaystyle\frac{p_{\theta}^2}{\mu r} - 1.
\end{equation}
Additionally, from Eq.~\eqref{eq:angular_momentum} the value of the eccentricity can be obtained:
\begin{equation}
    e^2 = e^2 \sin^2(\nu) + e^2 \cos^2(\nu) = 1 - \displaystyle\frac{p_{\theta}^2}{\mu a},
\end{equation}
which combined with Eqs.~\eqref{eq:semi-major} and~\eqref{eq:ecos} leads to:
\begin{equation} \label{eq:esin}
    e\sin(\nu) = e_x\sin(\theta) - e_y\cos(\theta) = \displaystyle\frac{p_{\theta}p_r}{\mu}.
\end{equation}
Using the relations from Eqs.~\eqref{eq:ecos} and~\eqref{eq:esin}, the orbital elements $e_x$ and $e_y$ can be obtained:
\begin{eqnarray}\label{eq:exey}
    e_x & = & \left(\displaystyle\frac{p_{\theta}^2}{\mu r} - 1\right)\cos(\theta) + \frac{p_{\theta}p_r}{\mu}\sin(\theta), \nonumber \\
    e_y & = & \left(\displaystyle\frac{p_{\theta}^2}{\mu r} - 1\right)\sin(\theta) - \frac{p_{\theta}p_r}{\mu}\cos(\theta).
\end{eqnarray}

The inclination of the orbit can be derived using the relations between the conjugate momenta $p_{\lambda}$ and the magnitude of the angular momentum $p_{\theta}$:
\begin{equation} \label{eq:inclination}
    i = \arccos\left(\displaystyle\frac{p_{\lambda}}{p_{\theta}}\right) = \arccos\left(\frac{p_{\lambda}\cos(\varphi)}{\sqrt{p_{\varphi}^2\cos^2(\varphi) + p_{\lambda}^2}}\right).
\end{equation}
The right ascension of the ascending node can be obtained using either spherical geometry, or as the constant of motion of the unperturbed problem resultant from the differential equation:
\begin{equation}
    \displaystyle\frac{d\varphi}{d\lambda} = \frac{p_{\varphi}}{p_{\lambda}}\cos^2(\varphi).
\end{equation}
In either case, the right ascension of the ascending node obtained is:
\begin{equation}\label{eq:rightascension}
\Omega = \begin{cases} \lambda - \arcsin\left(\sin(\varphi)\sqrt{\displaystyle\frac{p_{\lambda}^2}{p_{\varphi}^2\cos^2(\varphi) + p_{\lambda}^2\sin^2(\varphi)}}\right) &\mbox{if } p_{\varphi} \geq 0 \\
\lambda + \arcsin\left(\sin(\varphi)\sqrt{\displaystyle\frac{p_{\lambda}^2}{p_{\varphi}^2\cos^2(\varphi) + p_{\lambda}^2\sin^2(\varphi)}}\right) + \pi & \mbox{if } p_{\varphi} < 0 \end{cases}.
\end{equation}

And finally, the argument of latitude can be obtained using spherical trigonometry and the previous results, in particular:
\begin{equation}
\theta = \begin{cases} \arcsin\left(\sin(\varphi)\sqrt{\displaystyle\frac{p_{\varphi}^2\cos^2(\varphi) + p_{\lambda}^2}{p_{\varphi}^2\cos^2(\varphi) + p_{\lambda}^2\sin^2(\varphi)}}\right) &\mbox{if } p_{\varphi} \geq 0 \\
\pi -  \arcsin\left(\sin(\varphi)\sqrt{\displaystyle\frac{p_{\varphi}^2\cos^2(\varphi) + p_{\lambda}^2}{p_{\varphi}^2\cos^2(\varphi) + p_{\lambda}^2\sin^2(\varphi)}}\right) & \mbox{if } p_{\varphi} < 0 \end{cases},
\end{equation}
or alternatively for its use in Eq.~\eqref{eq:exey}:
\begin{eqnarray} \label{eq:sincostheta}
    \sin(\theta) & = & \displaystyle\frac{\sin(\varphi)}{\sin(i)} = \sin(\varphi)\sqrt{\displaystyle\frac{p_{\varphi}^2\cos^2(\varphi) + p_{\lambda}^2}{p_{\varphi}^2\cos^2(\varphi) + p_{\lambda}^2\sin^2(\varphi)}}, \nonumber \\
    \cos(\theta) & = & \displaystyle\frac{p_{\varphi}}{p_{\theta}}\frac{\cos(\varphi)}{\sin(i)} = \frac{p_{\varphi}\cos^2(\varphi)}{\sqrt{p_{\varphi}^2\cos^2(\varphi) + p_{\lambda}^2\sin^2(\varphi)}}.
\end{eqnarray}

\subsection{From orbital to spherical elements}

In the same way, it is also interesting to have a direct transformation from the orbital elements ($A$, $e_x$, $e_y$, $i$, $\Omega$, $\theta$) to spherical coordinates ($r$, $p_r$, $\varphi$, $p_{\varphi}$, $\lambda$, $p_{\lambda}$). This allows to perform all the substitutions required in the differential equations.

The magnitude of the angular momentum $p_{\theta}$ can be obtained using Eq.~\eqref{eq:A_def}:
\begin{equation}
    p_{\theta} = \sqrt[\leftroot{-1}\uproot{2}\scriptstyle 4]{\frac{R^2 \mu^2}{A}}.
\end{equation}
Another straight-forward transformation to derive is the radial distance. From Eq.~\eqref{eq:radial} the inverse of the radial distance is:
\begin{equation}
    \displaystyle\frac{1}{r} = \mu\frac{1 + e_x\cos(\theta) + e_y\sin(\theta)}{p_{\theta}^2}.
\end{equation}
Using these results and the system of equations from Eq.~\eqref{eq:exey}, it is possible to obtain the conjugate momenta of the radial distance $p_r$:
\begin{equation}
    p_r = \displaystyle\frac{\mu}{p_{\theta}}\left(e_x\sin(\theta) - e_y\cos(\theta)\right).
\end{equation}
The latitude of the orbit $\varphi$ can be derived using spherical trigonometry:
\begin{equation}
    \varphi = \arcsin\left(\sin(\theta)\sin(i)\right),
\end{equation}
and this result used in conjunction with Eq.\eqref{eq:sincostheta} to obtain the conjugate momenta of the latitude $p_{\varphi}$:
\begin{equation}
    p_{\varphi} = p_{\theta}\sin(i)\displaystyle\frac{\cos(\theta)}{\cos(\varphi)}.
\end{equation}
The latitude of the orbit can be obtained from Eq.~\eqref{eq:rightascension} by performing the substitution of all the spherical variables by the orbital elements:
\begin{equation}\label{eq:longitude}
\lambda = \begin{cases} \Omega + \arcsin\left(\displaystyle\frac{\sin(i)\cos(i)\sin(\theta)}{\sqrt{\cos^2(\theta) + \cos^2(i)\sin^2(i)\sin^2(\theta)}}\right) &\mbox{if } \cos(\theta) \geq 0 \\
\Omega + \pi - \arcsin\left(\displaystyle\frac{\sin(i)\cos(i)\sin(\theta)}{\sqrt{\cos^2(\theta) + \cos^2(i)\sin^2(i)\sin^2(\theta)}}\right) & \mbox{if } \cos(\theta) < 0 \end{cases}.
\end{equation}
Finally, the conjugate momenta of the longitude $p_{\lambda}$ is obtained using Eq.~\eqref{eq:inclination}:
\begin{equation}
    p_{\lambda} = p_{\theta}\cos(i).
\end{equation}

\subsection{Differential equation}

Once all the transformations are defined, it is possible to derive the system of differential equations for the orbital elements ($A$, $e_x$, $e_y$, $i$, $\Omega$, $\theta$):
\begin{eqnarray} \label{eq:difftemp}
	\displaystyle\frac{dA}{dt} & = & 12 J_2 \sqrt[\leftroot{-1}\uproot{2}\scriptstyle 4]{\frac{\mu^2 A^{11}}{R^6}} \left(1 + e_x\cos(\theta) + e_y\sin(\theta)\right)^3 \sin(\theta) \cos(\theta) \sin^2(i); \nonumber \\
	\displaystyle\frac{de_x}{dt} & = &  \frac{3}{2} J_2 \sqrt[\leftroot{-1}\uproot{2}\scriptstyle 4]{\frac{\mu^2 A^7}{R^6}} \sin(\theta) \left(1 + e_x\cos(\theta) + e_y\sin(\theta)\right)^3 \nonumber \\
	& \times & \big(-2e_y\cos^2(i)\sin(\theta) + (1 + e_x\cos(\theta) + e_y\sin(\theta))(3\sin^2(i)\sin^2(\theta) - 1) \nonumber \\
	& - & \sin^2(i)\cos(\theta)(3e_x + 4\cos(\theta) + e_x\cos(2\theta) + e_y\sin(2\theta)))\big); \nonumber \\
	\displaystyle\frac{de_y}{dt} & = & -\frac{3}{2} J_2 \sqrt[\leftroot{-1}\uproot{2}\scriptstyle 4]{\frac{\mu^2 A^7}{R^6}} \left(1 + e_x\cos(\theta) + e_y\sin(\theta)\right)^3 \nonumber \\
	& \times & \big(2e_y\cos^3(\theta)\sin^2(i)\sin(\theta) + e_x\cos^2(\theta)(5\sin^2(i)\sin^2(\theta) - 1) \nonumber \\
	& - &  2e_x\cos^2(i)\sin^2(\theta) + \cos(\theta)(1 + e_y\sin(\theta))(7\sin^2(i)\sin^2(\theta) - 1)\big); \nonumber \\
	\displaystyle\frac{di}{dt} & = & -3 J_2 \sqrt[\leftroot{-1}\uproot{2}\scriptstyle 4]{\frac{\mu^2 A^7}{R^6}} \left(1 + e_x\cos(\theta) + e_y\sin(\theta)\right)^3 \sin(i)\cos(i)\sin(\theta)\cos(\theta); \nonumber \\
	\displaystyle\frac{d\Omega}{dt} & = & -3 J_2 \sqrt[\leftroot{-1}\uproot{2}\scriptstyle 4]{\frac{\mu^2 A^7}{R^6}} \left(1 + e_x\cos(\theta) + e_y\sin(\theta)\right)^3 \cos(i) \sin^2(\theta); \nonumber \\
	\displaystyle\frac{d\theta}{dt} & = &  \sqrt[\leftroot{-1}\uproot{2}\scriptstyle 4]{\frac{\mu^2 A^3}{R^6}} \left(1 + e_x\cos(\theta) + e_y\sin(\theta)\right)^2 \nonumber \\
	& \times & \left(1 + 3 J_2 A \left(1 + e_x\cos(\theta) + e_y\sin(\theta)\right)\cos^2(i) \sin^2(\theta)\right).
\end{eqnarray}
This is still an exact differential equation, that is, no approximation has been performed up to this point. However, in order to provide an analytical solution to this system, even if it is an approximation, we require to perform an additional transformation to the independent variable $t$, and apply a perturbation method.


\section{Perturbation method}
\label{sec:perturbation}

This work uses a perturbation method based on the expansion of all the variables of the system as a power series of the small parameter $J_2$. In that sense, the approach is similar to the Poincar\'e-Lindstedt method, however, instead of perturbing the frequencies of the solution to cancel the secular effects on the orbital elements, in here a time regularization is used with the goal of matching the frequency of the solution with the variation of the perturbation. This is done by the use of the perturbed osculating value of the argument of latitude of the orbit. This approach allows to have a good representation of the system, especially close to the frozen conditions since the perturbed frequency on this region is closer to the unperturbed frequency. However, as the orbits get further away from the frozen condition, the accuracy of this approximation is slightly degraded due to the difference in the perturbed frequency compared with the unperturbed frequency of each orbital element. Nevertheless, this has the advantage of not having to evaluate the different frequencies of the solution as it was done in Arnas and Linares~\cite{zonal} while still providing a good accuracy in the solution for any kind of orbit. 

Particularly, the following time regularization is performed in the system using the argument of latitude of the orbit:
\begin{eqnarray}
    \displaystyle\frac{d\theta}{dt} & = &  \sqrt[\leftroot{-1}\uproot{2}\scriptstyle 4]{\frac{\mu^2 A^3}{R^6}} \left(1 + e_x\cos(\theta) + e_y\sin(\theta)\right)^2 \nonumber \\
	& \times & \left(1 + 3 J_2 A \left(1 + e_x\cos(\theta) + e_y\sin(\theta)\right)\cos^2(i) \sin^2(\theta)\right).
\end{eqnarray}
This time regularization is then introduced in the differential equation to obtain the following exact system:
\begin{eqnarray}
	\displaystyle\frac{dA}{d\theta} & = & 12 \frac{J_2 A^2}{\Delta} \left(1 + e_x\cos(\theta) + e_y\sin(\theta)\right) \sin(\theta) \cos(\theta) \sin^2(i); \nonumber \\
	\displaystyle\frac{de_x}{d\theta} & = &  \frac{3}{2} \frac{J_2 A}{\Delta} \sin(\theta) \left(1 + e_x\cos(\theta) + e_y\sin(\theta)\right) \nonumber \\
	& \times & \big(-2e_y\cos^2(i)\sin(\theta) + (1 + e_x\cos(\theta) + e_y\sin(\theta))(3\sin^2(i)\sin^2(\theta) - 1) \nonumber \\
	& - & \sin^2(i)\cos(\theta)(3e_x + 4\cos(\theta) + e_x\cos(2\theta) + e_y\sin(2\theta)))\big); \nonumber \\
	\displaystyle\frac{de_y}{d\theta} & = & -\frac{3}{2} \frac{J_2 A}{\Delta} \left(1 + e_x\cos(\theta) + e_y\sin(\theta)\right) \nonumber \\
	& \times & \big(2e_y\cos^3(\theta)\sin^2(i)\sin(\theta) + e_x\cos^2(\theta)(5\sin^2(i)\sin^2(\theta) - 1) \nonumber \\
	& - &  2e_x\cos^2(i)\sin^2(\theta) + \cos(\theta)(1 + e_y\sin(\theta))(7\sin^2(i)\sin^2(\theta) - 1)\big); \nonumber \\
	\displaystyle\frac{di}{d\theta} & = & -3 \frac{J_2 A}{\Delta} \left(1 + e_x\cos(\theta) + e_y\sin(\theta)\right) \sin(i)\cos(i)\sin(\theta)\cos(\theta); \nonumber \\
	\displaystyle\frac{d\Omega}{d\theta} & = & -3 \frac{J_2 A}{\Delta} \left(1 + e_x\cos(\theta) + e_y\sin(\theta)\right) \cos(i) \sin^2(\theta); \nonumber \\
	\displaystyle\frac{dt}{d\theta} & = &  \sqrt[\leftroot{-1}\uproot{2}\scriptstyle 4]{\frac{R^6}{\mu^2 A^3}}\frac{1}{\Delta \left(1 + e_x\cos(\theta) + e_y\sin(\theta)\right)^2}.
\end{eqnarray}
where:
\begin{equation}
    \Delta = 1 + 3 J_2 A \left(1 + e_x\cos(\theta) + e_y\sin(\theta)\right)\cos^2(i) \sin^2(\theta).
\end{equation}
This is the system of equations in which the perturbation method is applied.

Let $\vec{x} = \{A, e_x, e_y, i, \Omega\}$ be the set of orbital elements in study. Then, we can represent the previous system of differential equation as:
\begin{equation}
    \displaystyle\frac{d\vec{x}}{dt} = f(\vec{x}).
\end{equation}
In this system, a power series expansion based on $J_2$ in both the variables and the differential equation can be performed, leading to:
\begin{eqnarray} \label{eq:difexpansion}
    \vec{x} & \approx & \sum_{j=0}^{n} \vec{x}_j J_2^j; \nonumber \\
    f(\vec{x}) & \approx & \sum_{j=0}^{n} f_j(\vec{x}) J_2^j = \sum_{j=0}^{n} f_j\left(\sum_{k=0}^{n} \vec{x}_k J_2^k\right) J_2^j;
\end{eqnarray}
where $n$ is the maximum order of the power series expansion. This approximation can be introduced in the differential equation to obtain:
\begin{equation}
    \displaystyle\frac{d\vec{x}}{d\theta} \approx \sum_{j=0}^{n} \frac{d\vec{x}_j}{d\theta} J_2^j \approx \sum_{j=0}^{n} f_j(\vec{x}) J_2^j = \sum_{j=0}^{n} f_j\left(\sum_{k=0}^{n} \vec{x}_k J_2^k\right) J_2^j,
\end{equation}
where an identification of terms based on the powers of $J_2$ can be performed, leading to a system of ordinary differential equations that is solvable by direct integration of each individual differential equation of the system. Particularly, and for a second order solution, an expansion of the orbital elements can be performed such that:
\begin{eqnarray}
    A & \approx & A_0 + A_1 J_2 + A_2 J_2^2, \nonumber \\
    e_x & \approx & e_{x0} + e_{x1} J_2 + e_{x2} J_2^2, \nonumber \\
    e_y & \approx & e_{y0} + e_{y1} J_2 + e_{y2} J_2^2, \nonumber \\
    i & \approx & i_0 + i_1 J_2 + i_2 J_2^2, \nonumber \\
    \Omega & \approx & \Omega_0 + \Omega_1 J_2 + \Omega_2 J_2^2,
\end{eqnarray}
where the subindices $0$ represent the solution of the unperturbed problem, the subindeces $1$ the first order approximation, and the subindices $2$, the second order approximation. We can introduce these expansions in the differential equation to obtain a second order approximation of $f(\vec{x})$ for each of the orbital elements, being the resultant approximate differential equations for variable $A$:
\begin{eqnarray}
    \displaystyle\frac{dA}{d\theta} & \approx & J_2 \left(6 A_0^2 \sin^2(i_0) (1 + e_{x0} \cos(\theta) + e_{y0} \sin(\theta)) \sin(2\theta)\right) \nonumber \\
    & + & J_2^2 \big(12 A_0 \cos(\theta) \sin(i_0) \sin(\theta) \big(2 (A_0 i_1 \cos(i) + A_1 \sin(i)) \nonumber \\
    & \times & (1 + e_{x0} \cos(\theta)  + e_{y0} \sin(\theta)) - 
   3 A_0^2 \cos^2(i_0) \sin(i_0) \sin^2(\theta) \nonumber \\
   & \times & (1 + e_{x0} \cos(\theta)  + e_{y0} \sin(\theta))^2 + 
   A_0 \sin(i_0) (e_{x1} \cos(\theta)  + e_{y1} \sin(\theta))\big)\big),
\end{eqnarray}
for the $x$ component of the eccentricity vector:
\begin{eqnarray}
    \displaystyle\frac{de_{x}}{d\theta} & \approx & \frac{3}{2} J_2 A_0\sin(\theta)(1 + e_{x0} \cos(\theta) + e_{y0} \sin(\theta))  \big(  -2 e_{y0} \cos^2(i_0) \sin(\theta) \nonumber \\
    & + & (1 + e_{x0} \cos(\theta) + e_{y0} \sin(\theta)) (-1 + 3 \sin^2(i_0) \sin^2(\theta)) \nonumber \\
    & - & \cos(\theta) \sin^2(i_0) (3 e_{x0} + 4 \cos(\theta) + e_{x0} \cos(2 \theta) + e_{y0} \sin(2 \theta))\big) \nonumber \\
    & + & \frac{3}{2} J_2^2 \sin(\theta) \big( (A_1 (1 + e_{x0} \cos(\theta) + e_{y0} \sin(\theta)) \nonumber \\
    & - & 3 A_0^2 \cos^2(i_0) \sin^2(\theta) (1 + e_{x0} \cos(\theta) + e_{y0} \sin(\theta))^2 \nonumber \\
      & + & A_0 (e_{x1} \cos(\theta) + e_{y1} \sin(\theta))) (-2 e_{y0} \cos^2(i_0) \sin(\theta) \nonumber \\
      & + & (1 + e_{x0} \cos(\theta) + e_{y0} \sin(\theta)) (-1 + 3 \sin^2(i_0) \sin^2(\theta)) \nonumber \\
      & - & \cos(\theta) \sin^2(i_0) (3 e_{x0} + 4 \cos(\theta) + e_{x0} \cos(2 \theta) + e_{y0} \sin(2 \theta))) \nonumber \\
      & + & A_0 (1 + e_{x0} \cos(\theta) + e_{y0} \sin(\theta)) (-2 e_{y1} \cos^2(i_0) \sin(\theta) \nonumber \\
      & + & 4 i_1 e_{y0} \cos(i_0) \sin(i_0) \sin(\theta) + 
      3 i_1 \sin(2 i_0) \sin^2(\theta) \nonumber \\
      & \times & (1 + e_{x0} \cos(\theta) + e_{y0} \sin(\theta)) + (e_{x1} \cos(\theta) + 
         e_{y1} \sin(\theta)) \nonumber \\
      & \times & (-1 + 3 \sin^2(i_0) \sin^2(\theta)) - 
      \cos(\theta) \sin^2(i_0) \nonumber \\
      & \times & (3 e_{x1} + e_{x1} \cos(2 \theta) + e_{y1} \sin(2 \theta)))\big),
\end{eqnarray}
for the $y$ component of the eccentricity vector:
\begin{eqnarray}
    \displaystyle\frac{de_{y}}{d\theta} & \approx & -\frac{3}{2} J_2 A_0\sin(\theta)(1 + e_{x0} \cos(\theta) + e_{y0} \sin(\theta))  \big( 2 e_{y0} \cos^3(\theta) \sin^2(i_0) \sin(\theta) \nonumber \\
     & - & 2 e_{x0} \cos^2(i_0) \sin(\theta)^2 + e_{x0} \cos^2(\theta) (-1 + 5 \sin^2(i_0) \sin^2(\theta)) \nonumber \\
     & + & \cos(\theta) (1 + e_{y0} \sin(\theta)) (-1 + 7 \sin^2(i_0) \sin^2(\theta))\big) \nonumber \\
    & + & J_2^2 \big(3 \cos(i_0) \sin^2(\theta) (((A_1 e_{x0} + A_0 e_{x1}) \cos(i_0) - 2 A_0 i_1 e_{x0} \sin(i_0)) \nonumber \\
    & \times & (1 + e_{x0} \cos(\theta) + 
       e_{y0} \sin(\theta)) + A_0 e_{x0} \cos(i_0) (e_{x1} \cos(\theta) + e_{y1} \sin(\theta))) \nonumber \\
    & + & \frac{9}{2} A_0^2 \cos^2(i_0) \sin^2(\theta) (1 + e_{x0} \cos(\theta) + e_{y0} \sin(\theta))^2 \nonumber \\
    & \times & (2 e_{y0} \cos(\theta)^3 \sin^2(i_0) \sin(\theta) - 
    2 e_{x0} \cos^2(i_0) \sin^2(\theta) \nonumber \\
    & + & e_{x0} \cos(\theta)^2 (-1 + 5 \sin^2(i_0) \sin^2(\theta)) + 
    \cos(\theta) (1 + e_{y0} \sin(\theta)) \nonumber \\
    & \times & (-1 + 7 \sin^2(i_0) \sin^2(\theta))) - \frac{3}{2} \cos(\theta) ((A_1 + (A_1 e_{x0} + A_0 e_{x1}) \cos(\theta) \nonumber \\
    & + & (A_1 e_{y0} + A_0 e_{y1}) \sin(\theta)) (2 e_{y0} \cos^2(\theta) \sin^2(i_0) \sin(\theta) \nonumber \\
    & + & e_{x0} \cos(\theta) (-1 + 5 \sin^2(i_0) \sin^2(\theta)) + (1 + e_{y0} \sin(\theta)) \nonumber \\
    & \times & (-1 + 7 \sin^2(i_0) \sin^2(\theta))) + 
    A_0 (1 + e_{x0} \cos(\theta) + e_{y0} \sin(\theta)) \nonumber \\
    & \times & (2 \cos^2(\theta) \sin(i_0) (2 i_1 e_{y0} \cos(i_0) + e_{y1} \sin(i_0)) \sin(\theta) \nonumber \\
    & + & \cos(\theta) (-e_{x1} + 5 e_{x1} \sin^2(i_0) \sin^2(\theta) + 5 i_1 e_{x0} \sin(2 i_0) \sin^2(\theta)) \nonumber \\
    & + & \sin(\theta) (7 i_1 \sin(2 i_0) \sin(\theta) (1 + e_{y0} \sin(\theta)) + e_{y1} (7 \sin^2(i_0) \sin^2(\theta) - 1))))\big),
\end{eqnarray}
for the inclination:
\begin{eqnarray}
    \displaystyle\frac{di}{d\theta} & \approx & -3 J_2 A_0 \cos(i_0) \cos(\theta) \sin(i_0) \sin(\theta) \big(1 + e_{x0} \cos(\theta) + e_{y0} \sin(\theta)\big) \nonumber \\
    & - & 3 J_2^2 \cos(\theta) \sin(\theta)  \big(\frac{1}{2} (2 A_0 i_1 \cos(2 i_0) + A_1 \sin(2 i_0)) \nonumber \\
    & \times & (1 + e_{x0} \cos(\theta) + e_{y0} \sin(\theta)) - 3 A_0^2 \cos^3(i_0) \sin(i_0) \sin^2(\theta) \nonumber \\
    & \times & (1 + e_{x0} \cos(\theta) + e_{y0} \sin(\theta))^2 \nonumber \\
    & + & A_0 \cos(i_0) \sin(i_0) (e_{x1} \cos(\theta) + e_{y1} \sin(\theta))\big),
\end{eqnarray}
and for the right ascension of the ascending node:
\begin{eqnarray}
    \displaystyle\frac{d\Omega}{d\theta} & \approx & -3 J_2 A_0 \cos(i_0) \sin^2(\theta) (1 + e_{x0} \cos(\theta) + e_{y0} \sin(\theta))  \nonumber \\
    & - & 3 J_2^2 \sin^2(\theta) \big((A_1 \cos(i_0) - A_0 i_1 \sin(i_0)) (1 + e_{x0} \cos(\theta) + e_{y0} \sin(\theta)) \nonumber \\
    & - & 3 A_0^2 \cos^3(i_0) \sin^2(\theta) (1 + e_{x0} \cos(\theta) + e_{y0} \sin(\theta))^2 \nonumber \\
    & + & A_0 \cos(i_0) (e_{x1} \cos(\theta) + e_{y1} \sin(\theta))\big),
\end{eqnarray}
This is the system of differential equations that will be solved sequentially in the following section to create the second order approximate analytical solution to the system. Note that the variables used are the osculating variables and not mean variables, so they describe the actual motion of the orbiter. Later, by averaging these results, the analytical transformation from osculating to mean elements is derived.


\section{Approximate analytic solution}

As can be seen in the approximated differential equation, the derivatives of the variables depend on the solution of the unperturbed system and the first order solution (subindexes $0$ and $1$ respectively). Particularly, if we perform the expansion of the differential as seen in Eq.~\eqref{eq:difexpansion}, the first order solution depends on the result for the unperturbed system, while the second order solution depends on both the first order solution an the unperturbed system. This means that it is necessary to solve this system of differential sequentially, starting from the unperturbed system and moving up to a second order solution. The following subsections cover each one of these solution series in detail. 

\subsection{Unperturbed system}

The solutions to $A_0$, $e_{x0}$, $e_{y0}$, $i_0$, and $\Omega_0$ are given by the following system of equations:
\begin{eqnarray}
	\displaystyle\frac{dA_0}{d\theta} = 0; & & \displaystyle\frac{di_0}{d\theta} = 0; \nonumber \\
	\displaystyle\frac{de_{x0}}{d\theta} =  0; & & \displaystyle\frac{d\Omega_0}{d\theta} = 0; \nonumber \\
	\displaystyle\frac{de_{y0}}{d\theta} = 0; & &
\end{eqnarray}
where as can be seen, all these variables represent constants of motion of the unperturbed system. This means that they are also equal to the osculating initial values of these variables, and thus: $A_0 = A(t = 0)$, $e_{x0} = e_{x}(t = 0)$, $e_{y0} = e_{y}(t = 0)$, $i_0 = i(t = 0)$, and $\Omega_0 = \Omega(t = 0)$, or conversely, if taking $\theta$ as the independent variable: $A_0 = A(\theta = \theta_0)$, $e_{x0} = e_{x}(\theta = \theta_0)$, $e_{y0} = e_{y}(\theta = \theta_0)$, $i_0 = i(\theta = \theta_0)$, and $\Omega_0 = \Omega(\theta = \theta_0)$, where $\theta_0$ is the initial argument of latitude.

\subsection{First order solution}\label{sec:firstorder}

The first order solution is defined by the following differential equation:
\begin{eqnarray}
	\displaystyle\frac{dA_1}{d\theta} & = &  6 A_0^2 \sin^2(i_0) (1 + e_{x0} \cos(\theta) + e_{y0} \sin(\theta)) \sin(2\theta); \nonumber \\
	\displaystyle\frac{de_{x01}}{d\theta} & = &  \frac{3}{2} A_0 \sin(\theta) (1 + e_{x0} \cos(\theta) + e_{y0} \sin(\theta)) (-2 e_{y0} \cos^2(i_0) \sin(\theta) \nonumber \\
    & + & (1 + e_{x0} \cos(\theta) + e_{y0} \sin(\theta)) (-1 + 3 \sin^2(i_0) \sin^2(\theta)) \nonumber \\
    & - & \cos(\theta) \sin^2(i_0) (3 e_{x0} + 4 \cos(\theta) + e_{x0} \cos(2 \theta) + e_{y0} \sin(2 \theta))); \nonumber \\
	\displaystyle\frac{de_{y1}}{d\theta} & = & -\frac{3}{2} A_0 (1 + e_{x0} \cos(\theta) + e_{y0} \sin(\theta)) (2 e_{y0} \cos^3(\theta) \sin^2(i_0) \sin(\theta) \nonumber \\
     & - & 2 e_{x0} \cos^2(i_0) \sin(\theta)^2 + e_{x0} \cos^2(\theta) (-1 + 5 \sin^2(i_0) \sin^2(\theta)) \nonumber \\
     & + & \cos(\theta) (1 + e_{y0} \sin(\theta)) (-1 + 7 \sin^2(i_0) \sin^2(\theta))); \nonumber \\
	\displaystyle\frac{di_1}{d\theta} & = & -3 A_0 \cos(i_0) \cos(\theta) \sin(i_0) \sin(\theta) (1 + e_{x0} \cos(\theta) + e_{y0} \sin(\theta)); \nonumber \\
	\displaystyle\frac{d\Omega_1}{d\theta} & = & -3 A_0 \cos(i_0) \sin^2(\theta) (1 + e_{x0} \cos(\theta) + e_{y0} \sin(\theta)),
\end{eqnarray}
where at the initial instant $A_1 = e_{x1} = e_{y1} = i_1 = \Omega_1 =0$, that is, at the initial instant the perturbation has not acted yet into the variables of the system. It is important to note that this set of differential equations is uncoupled and thus, the evolution of the first order solution variables can be obtained by a direct integration of each individual differential equation. That way, the first order solution for variable $A$ is:
\begin{eqnarray}
A_1 & = & A^2 \sin ^2(i) \Bigl(4 e_x \cos ^3(\theta_0)-4 e_y \sin ^3(\theta_0)+3 \cos (2 \theta_0) \nonumber \\
& - & 4 e_x \cos ^3(\theta)+4 e_y \sin ^3(\theta)-3 \cos (2 \theta)\Bigl),
\end{eqnarray}
the first order solution of the $x$ component of the eccentricity vector is:
\begin{eqnarray}
e_{x1} & = & \frac{1}{128} A \Bigl(18 \cos (2 i-\theta_0) e_x^2-84 \cos (\theta_0) e_x^2 - 38 \cos (3 \theta_0) e_x^2 \nonumber \\
& - & 6 \cos (5 \theta_0) e_x^2+18 \cos (2 i+\theta_0) e_x^2+11 \cos (2 i+3 \theta_0) e_x^2 \nonumber \\
& + & 3 \cos (2 i+5 \theta_0) e_x^2+11 \cos (2 i-3 \theta_0) e_x^2-18 \cos (2 i-\theta) e_x^2 \nonumber \\
& + & 84 \cos (\theta) e_x^2+38 \cos (3 \theta) e_x^2+6 \cos (5 \theta) e_x^2-18 \cos (2 i+\theta) e_x^2 \nonumber \\ 
& - & 11 \cos (2 i+3 \theta) e_x^2-3 \cos (2 i+5 \theta) e_x^2-11 \cos (2 i-3 \theta) e_x^2 \nonumber \\
& + & 24 \cos (2 (i-\theta_0)) e_x-144 \cos (2 \theta_0) e_x-36 \cos (4 \theta_0) e_x \nonumber \\
& + & 24 \cos (2 (i+\theta_0)) e_x+18 \cos (2 (i+2 \theta_0)) e_x+18 \cos (2 (i-2 \theta_0)) e_x \nonumber \\
& - & 24 \cos (2 (i-\theta)) e_x+144 \cos (2 \theta) e_x+36 \cos (4 \theta) e_x \nonumber \\
& - & 24 \cos (2 (i+\theta)) e_x-18 \cos (2 (i+2 \theta)) e_x-18 \cos (2 (i-2 \theta)) e_x \nonumber \\
& - & 12 e_y \sin (2 i-\theta_0) e_x+168 e_y \sin (\theta_0) e_x-36 e_y \sin (3 \theta_0) e_x \nonumber \\
& - & 12 e_y \sin (5 \theta_0) e_x+12 e_y \sin (2 i+\theta_0) e_x-14 e_y \sin (2 i+3 \theta_0) e_x \nonumber \\
& + & 6 e_y \sin (2 i+5 \theta_0) e_x+14 e_y \sin (2 i-3 \theta_0) e_x-6 e_y \sin (2 i-5 \theta_0) e_x \nonumber \\
& + & 12 e_y \sin (2 i-\theta) e_x-168 e_y \sin (\theta) e_x+36 e_y \sin (3 \theta) e_x \nonumber \\
& + & 12 e_y \sin (5 \theta) e_x-12 e_y \sin (2 i+\theta) e_x+14 e_y \sin (2 i+3 \theta) e_x \nonumber \\
& - & 6 e_y \sin (2 i+5 \theta) e_x-14 e_y \sin (2 i-3 \theta) e_x+6 e_y \sin (2 i-5 \theta) e_x \nonumber \\
& + & 144 \theta_0 e_y-144 \theta e_y+240 \theta_0 e_y \cos (2 i)-240 \theta e_y \cos (2 i) \nonumber \\
& - & 150 e_y^2 \cos (2 i-\theta_0)-60 \cos (2 i-\theta_0)-132 e_y^2 \cos (\theta_0) \nonumber \\
& - & 72 \cos (\theta_0)-2 e_y^2 \cos (3 \theta_0)-56 \cos (3 \theta_0)+6 e_y^2 \cos (5 \theta_0) \nonumber \\
& - & 150 e_y^2 \cos (2 i+\theta_0)-60 \cos (2 i+\theta_0)+25 e_y^2 \cos (2 i+3 \theta_0) \nonumber \\
& + & 28 \cos (2 i+3 \theta_0)-3 e_y^2 \cos (2 i+5 \theta_0)+25 e_y^2 \cos (2 i-3 \theta_0) \nonumber \\
& + & 28 \cos (2 i-3 \theta_0)+3 \left(e_x^2-e_y^2\right) \cos (2 i-5 \theta_0)+150 e_y^2 \cos (2 i-\theta) \nonumber \\
& + & 60 \cos (2 i-\theta)+132 e_y^2 \cos (\theta)+72 \cos (\theta)+2 e_y^2 \cos (3 \theta)+56 \cos (3 \theta) \nonumber \\
& - & 6 e_y^2 \cos (5 \theta)+150 e_y^2 \cos (2 i+\theta)+60 \cos (2 i+\theta)-25 e_y^2 \cos (2 i+3 \theta) \nonumber \\
& - & 28 \cos (2 i+3 \theta)+3 e_y^2 \cos (2 i+5 \theta)-25 e_y^2 \cos (2 i-3 \theta) \nonumber \\
& - & 28 \cos (2 i-3 \theta)-3 \left(e_x^2-e_y^2\right) \cos (2 i-5 \theta)+96 e_y \sin (2 (i-\theta_0)) \nonumber \\
& - & 36 e_y \sin (4 \theta_0)-96 e_y \sin (2 (i+\theta_0))+18 e_y \sin (2 (i+2 \theta_0)) \nonumber \\
& - & 18 e_y \sin (2 (i-2 \theta_0))-96 e_y \sin (2 (i-\theta))+36 e_y \sin (4 \theta) \nonumber \\
& + & 96 e_y \sin (2 (i+\theta))-18 e_y \sin (2 (i+2 \theta))+18 e_y \sin (2 (i-2 \theta))\Bigl),
\end{eqnarray}
the first order solution of the $y$ component of the eccentricity vector is:
\begin{eqnarray}
e_{y1} & = & -\frac{1}{64} A \Bigl(6 \sin ^2(i) \sin (5 \theta_0) \left(e_x^2-e_y^2\right) \nonumber \\
& + & 6 \sin (\theta_0) \bigl(\cos (2 i) \left(9 e_x^2+9 e_y^2+14\right)+11 e_x^2-5 e_y^2+2\bigl) \nonumber \\
& - & \sin (3 \theta_0) \left(\cos (2 i) \left(13 e_x^2+23 e_y^2+28\right)-5 e_x^2-15 e_y^2-28\right) \nonumber \\
& - & 12 e_x e_y (13 \cos (2 i)+3) \cos (\theta_0)+20 e_x e_y \sin ^2(i) \cos (3 \theta_0) \nonumber \\
& - & 12 e_x e_y \sin ^2(i) \cos (5 \theta_0)+48 e_x \sin ^2(i) \sin (2 \theta_0)\nonumber \\
& + & 36 e_x \sin ^2(i) \sin (4 \theta_0) + 24 \theta_0 e_x (5 \cos (2 i)+3)\nonumber \\
& - & 48 e_y (2 \cos (2 i)-1) \cos (2 \theta_0) - 36 e_y \sin ^2(i) \cos (4 \theta_0)\nonumber \\
& + & 6 \sin ^2(i) \sin (5 \theta) \left(e_y^2-e_x^2\right)\nonumber \\
& - & 6 \sin (\theta) \left(\cos (2 i) \left(9 e_x^2+9 e_y^2+14\right)+11 e_x^2-5 e_y^2+2\right)\nonumber \\
& + & \sin (3 \theta) \left(\cos (2 i) \left(13 e_x^2+23 e_y^2+28\right)-5 e_x^2-15 e_y^2-28\right)\nonumber \\
& + & 12 e_x e_y (13 \cos (2 i)+3) \cos (\theta)-20 e_x e_y \sin ^2(i) \cos (3 \theta)\nonumber \\
& + & 12 e_x e_y \sin ^2(i) \cos (5 \theta)-48 e_x \sin ^2(i) \sin (2 \theta)\nonumber \\
& - & 36 e_x \sin ^2(i) \sin (4 \theta) - 24 \theta e_x (5 \cos (2 i)+3)\nonumber \\
& + & 48 e_y (2 \cos (2 i)-1) \cos (2 \theta)+36 e_y \sin ^2(i) \cos (4 \theta)\Bigl),
\end{eqnarray}
the first order solution of the inclination of the orbits is:
\begin{eqnarray}
i_1 & = & -\frac{1}{2} A \sin (i) \cos (i) \Bigl(2 e_y \left(\sin ^3(\theta)-\sin ^3(\theta_0)\right)+2 e_x \cos ^3(\theta_0)\nonumber \\
& + & 3 \cos ^2(\theta_0)-2 e_x \cos ^3(\theta)-3 \cos ^2(\theta)\Bigl),
\end{eqnarray}
and the first order solution of the right ascension of the ascending node is:
\begin{eqnarray}
\Omega_1 & = & \frac{1}{4} A \cos (i) \Bigl(4 e_x \left(\sin ^3(\theta_0)-\sin ^3(\theta)\right)\nonumber \\
& + & e_y (-9 \cos (\theta_0)+\cos (3 \theta_0)+9 \cos (\theta)-\cos (3 \theta))\nonumber \\
& + & 6 (\theta_0-\sin (\theta_0) \cos (\theta_0)-\theta+\sin (\theta) \cos (\theta))\Bigl),
\end{eqnarray}
where $\theta$ represents the final argument of latitude of the orbit.

These set of solutions already provides a relatively good accuracy for orbits ranging from circular to hyperbolic as seen in Section~\ref{sec:examples}. Nevertheless, due to the simple nature of these solutions and the methodology proposed, it is possible to extend the solution to a second order approximation which significantly improves the accuracy of the solution. 
	

\subsection{Second order solution} \label{sec:secondorder}

As performed in the previous section, we can define the differential equation related with the second order approximation by using the terms in power $J_2^2$ from the approximated differential equations. That way, the following system of equations is derived:
\begin{eqnarray}
	\displaystyle\frac{dA_2}{d\theta} & = & 12 A_0 \cos(\theta) \sin(i_0) \sin(\theta) \big(2 (A_0 i_1 \cos(i) + A_1 \sin(i)) \nonumber \\
    & \times & (1 + e_{x0} \cos(\theta)  + e_{y0} \sin(\theta)) - 
   3 A_0^2 \cos^2(i_0) \sin(i_0) \sin^2(\theta) \nonumber \\
   & \times & (1 + e_{x0} \cos(\theta)  + e_{y0} \sin(\theta))^2 + 
   A_0 \sin(i_0) (e_{x1} \cos(\theta)  + e_{y1} \sin(\theta))\big); \nonumber \\
	\displaystyle\frac{de_{x2}}{d\theta} & = &  \frac{3}{2} \sin(\theta) ((A_1 (1 + e_{x0} \cos(\theta) + e_{y0} \sin(\theta)) \nonumber \\
    & - & 3 A_0^2 \cos^2(i_0) \sin^2(\theta) (1 + e_{x0} \cos(\theta) + e_{y0} \sin(\theta))^2 \nonumber \\
      & + & A_0 (e_{x1} \cos(\theta) + e_{y1} \sin(\theta))) (-2 e_{y0} \cos^2(i_0) \sin(\theta) \nonumber \\
      & + & (1 + e_{x0} \cos(\theta) + e_{y0} \sin(\theta)) (-1 + 3 \sin^2(i_0) \sin^2(\theta)) \nonumber \\
      & - & \cos(\theta) \sin^2(i_0) (3 e_{x0} + 4 \cos(\theta) + e_{x0} \cos(2 \theta) + e_{y0} \sin(2 \theta))) \nonumber \\
      & + & A_0 (1 + e_{x0} \cos(\theta) + e_{y0} \sin(\theta)) (-2 e_{y1} \cos^2(i_0) \sin(\theta) \nonumber \\
      & + & 4 i_1 e_{y0} \cos(i_0) \sin(i_0) \sin(\theta) + 
      3 i_1 \sin(2 i_0) \sin^2(\theta) \nonumber \\
      & \times & (1 + e_{x0} \cos(\theta) + e_{y0} \sin(\theta)) + (e_{x1} \cos(\theta) + 
         e_{y1} \sin(\theta)) \nonumber \\
      & \times & (-1 + 3 \sin^2(i_0) \sin^2(\theta)) - 
      \cos(\theta) \sin^2(i_0) \nonumber \\
      & \times & (3 e_{x1} + e_{x1} \cos(2 \theta) + e_{y1} \sin(2 \theta)))); \nonumber \\
	\displaystyle\frac{de_{y2}}{d\theta} & = & 3 \cos(i_0) \sin^2(\theta) (((A_1 e_{x0} + A_0 e_{x1}) \cos(i_0) - 2 A_0 i_1 e_{x0} \sin(i_0)) \nonumber \\
    & \times & (1 + e_{x0} \cos(\theta) + 
       e_{y0} \sin(\theta)) + A_0 e_{x0} \cos(i_0) (e_{x1} \cos(\theta) + e_{y1} \sin(\theta))) \nonumber \\
    & + & \frac{9}{2} A_0^2 \cos^2(i_0) \sin^2(\theta) (1 + e_{x0} \cos(\theta) + e_{y0} \sin(\theta))^2 \nonumber \\
    & \times & (2 e_{y0} \cos(\theta)^3 \sin^2(i_0) \sin(\theta) - 
    2 e_{x0} \cos^2(i_0) \sin^2(\theta) \nonumber \\
    & + & e_{x0} \cos(\theta)^2 (-1 + 5 \sin^2(i_0) \sin^2(\theta)) + 
    \cos(\theta) (1 + e_{y0} \sin(\theta)) \nonumber \\
    & \times & (-1 + 7 \sin^2(i_0) \sin^2(\theta))) - \frac{3}{2} \cos(\theta) ((A_1 + (A_1 e_{x0} + A_0 e_{x1}) \cos(\theta) \nonumber \\
    & + & (A_1 e_{y0} + A_0 e_{y1}) \sin(\theta)) (2 e_{y0} \cos^2(\theta) \sin^2(i_0) \sin(\theta) \nonumber \\
    & + & e_{x0} \cos(\theta) (-1 + 5 \sin^2(i_0) \sin^2(\theta)) + (1 + e_{y0} \sin(\theta)) \nonumber \\
    & \times & (-1 + 
          7 \sin^2(i_0) \sin^2(\theta))) + 
    A_0 (1 + e_{x0} \cos(\theta) + 
       e_{y0} \sin(\theta)) \nonumber \\
    & \times & (2 \cos^2(\theta) \sin(i_0) (2 i_1 e_{y0} \cos(i_0) + e_{y1} \sin(i_0)) \sin(\theta) \nonumber \\
    & + & \cos(\theta) (-e_{x1} + 5 e_{x1} \sin^2(i_0) \sin^2(\theta) + 
          5 i_1 e_{x0} \sin(2 i_0) \sin^2(\theta)) \nonumber \\
    & + & \sin(\theta) (7 i_1 \sin(2 i_0) \sin(\theta) (1 + e_{y0} \sin(\theta)) + 
          e_{y1} (7 \sin^2(i_0) \sin^2(\theta) - 1)))); \nonumber \\
	\displaystyle\frac{di_2}{d\theta} & = & -3 \cos(\theta) \sin(\theta) (\frac{1}{2} (2 A_0 i_1 \cos(2 i_0) + A_1 \sin(2 i_0)) (1 + e_{x0} \cos(\theta) + e_{y0} \sin(\theta)) \nonumber \\
   & - & 3 A_0^2 \cos^3(i_0) \sin(i_0) \sin^2(\theta) (1 + e_{x0} \cos(\theta) + e_{y0} \sin(\theta))^2 \nonumber \\
   & + & A_0 \cos(i_0) \sin(i_0) (e_{x1} \cos(\theta) + e_{y1} \sin(\theta))); \nonumber \\
	\displaystyle\frac{d\Omega_2}{d\theta} & = & -3 \sin^2(\theta) ((A_1 \cos(i_0) - A_0 i_1 \sin(i_0)) (1 + e_{x0} \cos(\theta) + e_{y0} \sin(\theta)) \nonumber \\
    & - & 3 A_0^2 \cos^3(i_0) \sin^2(\theta) (1 + e_{x0} \cos(\theta) + e_{y0} \sin(\theta))^2 \nonumber \\
    & + & A_0 \cos(i_0) (e_{x1} \cos(\theta) + e_{y1} \sin(\theta))),
\end{eqnarray}
where, as before, $A_2 = e_{x2} = e_{y2} = i_2 = \Omega_2 = 0$ at the initial instant. This is again a system of differential equations that is uncoupled and thus, can be solved by direct integration of each differential equation of the system. One thing to note is that the differential equation contains terms in the first order solution. These terms must be substituted by the results obtained in the first order solution before performing the integral. As expected, the solution becomes relatively large, so its expression is not included in this manuscript. Instead, the Matlab code containing these expressions as well as all the transformations from osculating to mean elements can be found in the web page: \href{https://engineering.purdue.edu/ART/research/research-code}{https://engineering.purdue.edu/ART/research/research-code}.

One important characteristic of this second order solution is that, as in the case of the first order solution, it only contains combinations of monomials with sines and cosines. This means that the analytical solution does not present singularities for any initial condition as opposed to other perturbation methods based on averaging where singularities can appear related with frozen orbits. 


\section{Transformation to mean elements}

The previous analytical approximations provide the evolution of the osculating orbital elements as a function of the argument of latitude of the orbits. Therefore, it is possible to use these results to obtain the mean motion of the system by averaging the equation for a single revolution. In that regard, having chosen the argument of latitude as the independent variable is going to perfectly define the limits of the integral for the purpose of averaging. Particularly, we assume that the orbiting object will have the following state vector $\vec{x} = \{A, \ e_x, \ e_y, \ i, \ \Omega\}$, so the mean state vector will be defined as:
\begin{equation}
    \overline{\vec{x}} = \displaystyle\frac{1}{2\pi}\int_{\theta_0 - \pi}^{\theta_0 + \pi} \vec{x} d\theta
\end{equation}
where the limits of integration have been selected such that the current state is centered on the domain of integration. Additionally, since the zero, first, and second order solutions are already written in terms of the initial state, it is possible to obtain the mean values of the orbital elements by averaging each particular order of the solution, that is:
\begin{eqnarray}
    \overline{A} & \approx & \overline{A_0} + \overline{A_1} J_2 + \overline{A_2} J_2^2, \nonumber \\
    \overline{e_x} & \approx & \overline{e_{x0}} + \overline{e_{x1}} J_2 + \overline{e_{x2}} J_2^2, \nonumber \\
    \overline{e_y} & \approx & \overline{e_{y0}} + \overline{e_{y1}} J_2 + \overline{e_{y2}} J_2^2, \nonumber \\
    \overline{i} & \approx & \overline{i_0} + \overline{i_1} J_2 + \overline{i_2} J_2^2, \nonumber \\
    \overline{\Omega} & \approx & \overline{\Omega_0} + \overline{\Omega_1} J_2 + \overline{\Omega_2} J_2^2.
\end{eqnarray}
In the following subsections, each order of the solution is covered independently.

\subsection{Zero order transformation}

The zero order transformation corresponds to the unperturbed Keplerian system, and as such, performing the averaging leads to: $\overline{A_0} = A(\theta_0)$, $\overline{e_{x0}} = e_{x}(\theta_0)$, $\overline{e_{y0}} = e_{y}(\theta_0)$, $\overline{i_0} = i(\theta_0)$, and $\overline{\Omega_0} = \Omega(\theta_0)$. In other words, the average zero order is equal to the osculating values of the orbital elements.

\subsection{First order transformation}

The first order transformation can be obtained by integration of the approximate solutions provided in Section~\ref{sec:firstorder}. Particularly, the result of these integrals lead to the following transformations:
\begin{eqnarray}
\overline{A_1} & = & A^2 \sin ^2(i) \Bigl(4 e_x \cos ^3(\theta_0)-4 e_y \sin ^3(\theta_0)+3 \cos (2 \theta_0)\Bigl); \nonumber \\
\overline{e_{x1}} & = & \frac{1}{128} A \Bigl(\left(11 e_x^2+25 e_y^2+28\right) \cos (2 i-3 \theta_0) + 3 \left(e_x^2-e_y^2\right) \cos (2 i-5 \theta_0)\nonumber \\
& + & 18 e_x^2 \cos (2 i-\theta_0)+18 e_x^2 \cos (2 i+\theta_0)+11 e_x^2 \cos (2 i+3 \theta_0)\nonumber \\
& + & 3 e_x^2 \cos (2 i+5 \theta_0)-12 e_x e_y \sin (2 i-\theta_0)+12 e_x e_y \sin (2 i+\theta_0)\nonumber \\
& - & 14 e_x e_y \sin (2 i+3 \theta_0)+6 e_x e_y \sin (2 i+5 \theta_0)+14 e_x e_y \sin (2 i-3 \theta_0)\nonumber \\
& - & 6 e_x e_y \sin (2 i-5 \theta_0)+24 e_x \cos (2 (i-\theta_0))+24 e_x \cos (2 (i+\theta_0))\nonumber \\
& + & 18 e_x \cos (2 (i+2 \theta_0))+18 e_x \cos (2 (i-2 \theta_0))-150 e_y^2 \cos (2 i-\theta_0)\nonumber \\
& - & 150 e_y^2 \cos (2 i+\theta_0)+25 e_y^2 \cos (2 i+3 \theta_0)-3 e_y^2 \cos (2 i+5 \theta_0)\nonumber \\
& + & 96 e_y \sin (2 (i-\theta_0))-96 e_y \sin (2 (i+\theta_0))+18 e_y \sin (2 (i+2 \theta_0))\nonumber \\
& - & 18 e_y \sin (2 (i-2 \theta_0))-60 \cos (2 i-\theta_0)-60 \cos (2 i+\theta_0)\nonumber \\
& + & 28 \cos (2 i+3 \theta_0)-84 e_x^2 \cos (\theta_0)-38 e_x^2 \cos (3 \theta_0)-6 e_x^2 \cos (5 \theta_0)\nonumber \\
& + & 168 e_x e_y \sin (\theta_0)-36 e_x e_y \sin (3 \theta_0)-12 e_x e_y \sin (5 \theta_0)\nonumber \\
& - & 144 e_x \cos (2 \theta_0)-36 e_x \cos (4 \theta_0)-132 e_y^2 \cos (\theta_0)-2 e_y^2 \cos (3 \theta_0)\nonumber \\
& + & 6 e_y^2 \cos (5 \theta_0)-36 e_y \sin (4 \theta_0)-72 \cos (\theta_0)-56 \cos (3 \theta_0)\Bigl); \nonumber \\
\overline{e_{y1}} & = & -\frac{1}{64} A \Bigl(6 \sin (\theta_0) \left(\cos (2 i) \left(9 e_x^2+9 e_y^2+14\right)+11 e_x^2-5 e_y^2+2\right)\nonumber \\
& + & \sin (3 \theta_0) \left(-\cos (2 i) \left(13 e_x^2+23 e_y^2+28\right)+5 e_x^2+15 e_y^2+28\right)\nonumber \\
& + & 6 \sin ^2(i) \sin (5 \theta_0) (e_x-e_y) (e_x+e_y)-12 e_x e_y (13 \cos (2 i)+3) \cos (\theta_0)\nonumber \\
& + & 20 e_x e_y \sin ^2(i) \cos (3 \theta_0)-12 e_x e_y \sin ^2(i) \cos (5 \theta_0)\nonumber \\
& + & 48 e_x \sin ^2(i) \sin (2 \theta_0)+36 e_x \sin ^2(i) \sin (4 \theta_0)\nonumber \\
& + & 48 e_y (1-2 \cos (2 i)) \cos (2 \theta_0)-36 e_y \sin ^2(i) \cos (4 \theta_0)\Bigl); \nonumber \\
\overline{i_1} & = &\frac{1}{4} A \sin (i) \cos (i) \Bigl(-4 e_x \cos ^3(\theta_0)+4 e_y \sin ^3(\theta_0)-6 \cos ^2(\theta_0)+3\Bigl); \nonumber \\
\overline{\Omega_1} & = & \frac{1}{4} A \cos (i) \Bigl(4 e_x \sin ^3(\theta_0)+e_y \cos (3 \theta_0)-3 \cos (\theta_0) (2 \sin (\theta_0)+3 e_y)\Bigl).
\end{eqnarray}


\subsection{Second order transformation}

Similarly, the second order transformation can be obtained by integration of the approximate solutions from Section~\ref{sec:secondorder}. Particularly, the transformation for the second order orbital element $A_2$ is:
\begin{eqnarray}
\overline{A_2} & = & -\frac{1}{4480} A^3 \sin ^2(i) \Bigl(12915 e_x^2 \cos (2 (i-\theta_0))+12915 e_x^2 \cos (2 (i+\theta_0))\nonumber \\
& + & 2940 e_x^2 \cos (2 (i+2 \theta_0))+245 e_x^2 \cos (2 (i+3 \theta_0))+2940 e_x^2 \cos (2 (i-2 \theta_0))\nonumber \\
& + & 245 e_x^2 \cos (2 (i-3 \theta_0))-1890 e_x e_y \sin (2 (i-\theta_0))+1890 e_x e_y \sin (2 (i+\theta_0))\nonumber \\
& + & 1470 e_x e_y \sin (2 (i+2 \theta_0))+490 e_x e_y \sin (2 (i+3 \theta_0))\nonumber \\
& - & 1470 e_x e_y \sin (2 (i-2 \theta_0))-490 e_x e_y \sin (2 (i-3 \theta_0))\nonumber \\
& + & 26040 e_x \cos (2 i-\theta_0)+26985 e_x \cos (2 i+\theta_0)+12915 e_x \cos (2 i+3 \theta_0)\nonumber \\
& + & 1491 e_x \cos (2 i+5 \theta_0)-135 e_x \cos (2 i+7 \theta_0)+12600 e_x \cos (2 i-3 \theta_0)\nonumber \\
& + & 1680 e_x \cos (2 i-5 \theta_0)-3675 e_y^2 \cos (2 (i-\theta_0))-3675 e_y^2 \cos (2 (i+\theta_0))\nonumber \\
& + & 1470 e_y^2 \cos (2 (i+2 \theta_0))-245 e_y^2 \cos (2 (i+3 \theta_0))+1470 e_y^2 \cos (2 (i-2 \theta_0))\nonumber \\
& - & 245 e_y^2 \cos (2 (i-3 \theta_0))-10920 e_y \sin (2 i-\theta_0)+10920 e_y \sin (2 i+\theta_0)\nonumber \\
& - & 840 e_y \sin (2 i+3 \theta_0)+1680 e_y \sin (2 i+5 \theta_0)+840 e_y \sin (2 i-3 \theta_0)\nonumber \\
& - & 1680 e_y \sin (2 i-5 \theta_0)+7560 \cos (2 (i-\theta_0))+7560 \cos (2 (i+\theta_0))\nonumber \\
& + & 2730 \cos (2 (i+2 \theta_0))+2730 \cos (2 (i-2 \theta_0))\nonumber \\
& + & 560 \cos (2 i) \left(35 e_x^2+65 e_y^2+37\right)-2310 e_x^2 \cos (2 \theta_0)-3360 e_x^2 \cos (4 \theta_0)\nonumber \\
& - & 490 e_x^2 \cos (6 \theta_0)+9660 e_x e_y \sin (2 \theta_0)+2100 e_x e_y \sin (4 \theta_0)\nonumber \\
& - & 980 e_x e_y \sin (6 \theta_0)+8400 e_x \cos (\theta_0)-5040 e_x \cos (3 \theta_0)-3360 e_x \cos (5 \theta_0)\nonumber \\
& + & 17430 e_y^2 \cos (2 \theta_0)-5460 e_y^2 \cos (4 \theta_0)+490 e_y^2 \cos (6 \theta_0)-28560 e_y \sin (\theta_0)\nonumber \\
& + & 21840 e_y \sin (3 \theta_0)-3360 e_y \sin (5 \theta_0)+11760 \cos (2 \theta_0)-5460 \cos (4 \theta_0)\nonumber \\
& - & 18480 e_x^2-6160 e_y^2-10640\Bigl),
\end{eqnarray}
for the second order $x$ component of the eccentricity vector $e_{x2}$ is:
\begin{eqnarray}
\overline{e_{x2}} & = & -\displaystyle\frac{1}{81920} A^2 (32200 e_x + 41280 \pi^2 e_x + 62800 e_x^3 + 
   260240 e_x e_y^2 \nonumber \\
& + & 
   160 e_x (-63 + 360 \pi^2 + 169 e_x^2 + 1549 e_y^2) \cos(2 i) \nonumber \\
& + & 
   40 e_x (23 + 600 \pi^2 + 378 e_x^2 + 3618 e_y^2) \cos(4 i) + 
   2520 e_x^2 \cos(2 i - 7 \theta_0) \nonumber \\
& - & 2520 e_y^2 \cos(2 i - 7 \theta_0) - 
   1170 e_x^2 \cos(4 i - 7 \theta_0) + 1170 e_y^2 \cos(4 i - 7 \theta_0) \nonumber \\
& - & 
   3360 e_x \cos(4 i - 6 \theta_0) - 200 e_x^3 \cos(4 i - 6 \theta_0) - 
   3960 e_x e_y^2 \cos(4 i - 6 \theta_0) \nonumber \\
& + & 6240 \cos(2 i - 5 \theta_0) + 
   9984 e_x^2 \cos(2 i - 5 \theta_0) + 9696 e_y^2 \cos(2 i - 5 \theta_0) \nonumber \\
& - & 
   3240 \cos(4 i - 5 \theta_0) - 180 e_x^2 \cos(4 i - 5 \theta_0) - 
   12420 e_y^2 \cos(4 i - 5 \theta_0) \nonumber \\
& + & 300 e_x^3 \cos(2 (i - 4 \theta_0)) - 
   900 e_x e_y^2 \cos(2 (i - 4 \theta_0)) + 6880 e_x \cos(2 (i - 3 \theta_0)) \nonumber \\
& + & 
   1920 e_x^3 \cos(2 (i - 3 \theta_0)) + 2240 e_x e_y^2 \cos(2 (i - 3 \theta_0)) + 
   18880 \cos(2 i - 3 \theta_0) \nonumber \\
& + & 2480 e_x^2 \cos(2 i - 3 \theta_0) + 
   65680 e_y^2 \cos(2 i - 3 \theta_0) - 5120 \cos(4 i - 3 \theta_0) \nonumber \\
& + & 
   6400 e_x^2 \cos(4 i - 3 \theta_0) + 22400 e_y^2 \cos(4 i - 3 \theta_0) + 
   23760 e_x \cos(2 (i - 2 \theta_0)) \nonumber \\
& + & 4560 e_x^3 \cos(2 (i - 2 \theta_0)) + 
   16080 e_x e_y^2 \cos(2 (i - 2 \theta_0)) - 135 e_x^3 \cos(4 (i - 2 \theta_0)) \nonumber \\
& + & 
   405 e_x e_y^2 \cos(4 (i - 2 \theta_0)) + 23760 e_x \cos(4 i - 2 \theta_0) + 
   3720 e_x^3 \cos(4 i - 2 \theta_0) \nonumber \\
& + & 12600 e_x e_y^2 \cos(4 i - 2 \theta_0) + 
   55200 e_x \cos(2 (i - \theta_0)) - 24960 e_x^3 \cos(2 (i - \theta_0)) \nonumber \\
& + & 
   83520 e_x e_y^2 \cos(2 (i - \theta_0)) - 5460 e_x \cos(4 (i - \theta_0)) + 
   1980 e_x^3 \cos(4 (i - \theta_0)) \nonumber
\end{eqnarray}
\begin{eqnarray}
& - & 11220 e_x e_y^2 \cos(4 (i - \theta_0)) + 
   75360 \cos(2 i - \theta_0) + 113400 e_x^2 \cos(2 i - \theta_0) \nonumber \\
& - & 
   95640 e_y^2 \cos(2 i - \theta_0) - 5400 \cos(4 i - \theta_0) + 
   54150 e_x^2 \cos(4 i - \theta_0) \nonumber \\
& - & 194190 e_y^2 \cos(4 i - \theta_0) + 
   90480 \cos(\theta_0) + 94980 e_x^2 \cos(\theta_0) \nonumber \\
& - & 50100 e_y^2 \cos(\theta_0) + 
   26400 e_x \cos(2 \theta_0) - 34320 e_x^3 \cos(2 \theta_0) \nonumber \\
& + & 107280 e_x e_y^2 \cos(2 \theta_0) + 
   18560 \cos(3 \theta_0) - 31840 e_x^2 \cos(3 \theta_0) \nonumber \\
& + & 132320 e_y^2 \cos(3 \theta_0) - 
   9720 e_x \cos(4 \theta_0) - 1560 e_x^3 \cos(4 \theta_0) + 24840 e_x e_y^2 \cos(4 \theta_0) \nonumber \\
& - & 
   6000 \cos(5 \theta_0) - 15768 e_x^2 \cos(5 \theta_0) - 21432 e_y^2 \cos(5 \theta_0) - 
   7040 e_x \cos(6 \theta_0) \nonumber \\
& - & 3440 e_x^3 \cos(6 \theta_0) - 4240 e_x e_y^2 \cos(6 \theta_0) - 
   2700 e_x^2 \cos(7 \theta_0) + 2700 e_y^2 \cos(7 \theta_0) \nonumber \\
& - & 330 e_x^3 \cos(8 \theta_0) + 
   990 e_x e_y^2 \cos(8 \theta_0) + 55200 e_x \cos(2 (i + \theta_0)) \nonumber \\
& - & 
   24960 e_x^3 \cos(2 (i + \theta_0)) + 83520 e_x e_y^2 \cos(2 (i + \theta_0)) - 
   5460 e_x \cos(4 (i + \theta_0)) \nonumber \\
& + & 1980 e_x^3 \cos(4 (i + \theta_0)) - 
   11220 e_x e_y^2 \cos(4 (i + \theta_0)) + 75360 \cos(2 i + \theta_0) \nonumber \\
& + & 
   113400 e_x^2 \cos(2 i + \theta_0) - 95640 e_y^2 \cos(2 i + \theta_0) + 
   23760 e_x \cos(2 (2 i + \theta_0)) \nonumber \\
& + & 3720 e_x^3 \cos(2 (2 i + \theta_0)) + 
   12600 e_x e_y^2 \cos(2 (2 i + \theta_0)) - 5400 \cos(4 i + \theta_0) \nonumber \\
& + & 
   54150 e_x^2 \cos(4 i + \theta_0) - 194190 e_y^2 \cos(4 i + \theta_0) + 
   23760 e_x \cos(2 (i + 2 \theta_0)) \nonumber \\
& + & 4560 e_x^3 \cos(2 (i + 2 \theta_0)) + 
   16080 e_x e_y^2 \cos(2 (i + 2 \theta_0)) - 135 e_x^3 \cos(4 (i + 2 \theta_0)) \nonumber \\
& + & 
   405 e_x e_y^2 \cos(4 (i + 2 \theta_0)) + 6880 e_x \cos(2 (i + 3 \theta_0)) + 
   1920 e_x^3 \cos(2 (i + 3 \theta_0)) \nonumber \\
& + & 2240 e_x e_y^2 \cos(2 (i + 3 \theta_0)) + 
   18880 \cos(2 i + 3 \theta_0) + 2480 e_x^2 \cos(2 i + 3 \theta_0) \nonumber \\
& + & 
   65680 e_y^2 \cos(2 i + 3 \theta_0) - 5120 \cos(4 i + 3 \theta_0) + 
   6400 e_x^2 \cos(4 i + 3 \theta_0) \nonumber \\
& + & 22400 e_y^2 \cos(4 i + 3 \theta_0) + 
   300 e_x^3 \cos(2 (i + 4 \theta_0)) - 900 e_x e_y^2 \cos(2 (i + 4 \theta_0)) \nonumber \\
& + & 
   6240 \cos(2 i + 5 \theta_0) + 9984 e_x^2 \cos(2 i + 5 \theta_0) + 
   9696 e_y^2 \cos(2 i + 5 \theta_0) \nonumber \\
& - & 3240 \cos(4 i + 5 \theta_0) - 
   180 e_x^2 \cos(4 i + 5 \theta_0) - 12420 e_y^2 \cos(4 i + 5 \theta_0) \nonumber \\
& - & 
   3360 e_x \cos(4 i + 6 \theta_0) - 200 e_x^3 \cos(4 i + 6 \theta_0) - 
   3960 e_x e_y^2 \cos(4 i + 6 \theta_0) \nonumber \\
& + & 2520 e_x^2 \cos(2 i + 7 \theta_0) - 
   2520 e_y^2 \cos(2 i + 7 \theta_0) - 1170 e_x^2 \cos(4 i + 7 \theta_0) \nonumber \\
& + & 
   1170 e_y^2 \cos(4 i + 7 \theta_0) - 5040 e_x e_y \sin(2 i - 7 \theta_0) + 
   2340 e_x e_y \sin(4 i - 7 \theta_0) \nonumber \\
& + & 3360 e_y \sin(4 i - 6 \theta_0) - 
   1680 e_x^2 e_y \sin(4 i - 6 \theta_0) + 2080 e_y^3 \sin(4 i - 6 \theta_0) \nonumber \\
& - & 
   288 e_x e_y \sin(2 i - 5 \theta_0) - 12240 e_x e_y \sin(4 i - 5 \theta_0) - 
   900 e_x^2 e_y \sin(2 (i - 4 \theta_0)) \nonumber \\
& + & 300 e_y^3 \sin(2 (i - 4 \theta_0)) - 
   6880 e_y \sin(2 (i - 3 \theta_0)) - 1760 e_x^2 e_y \sin(2 (i - 3 \theta_0)) \nonumber \\
& - & 
   2080 e_y^3 \sin(2 (i - 3 \theta_0)) + 69440 e_x e_y \sin(2 i - 3 \theta_0) - 
   11240 e_x e_y \sin(4 i - 3 \theta_0) \nonumber \\
& + & 1200 e_y \sin(2 (i - 2 \theta_0)) + 
   2880 e_x^2 e_y \sin(2 (i - 2 \theta_0)) - 8640 e_y^3 \sin(2 (i - 2 \theta_0)) \nonumber \\
& + & 
   405 e_x^2 e_y \sin(4 (i - 2 \theta_0)) - 135 e_y^3 \sin(4 (i - 2 \theta_0)) + 
   72000 e_y \sin(4 i - 2 \theta_0) \nonumber \\
& + & 8880 e_x^2 e_y \sin(4 i - 2 \theta_0) + 
   12960 e_y^3 \sin(4 i - 2 \theta_0) + 51360 e_y \sin(2 (i - \theta_0)) \nonumber \\
& + & 
   97440 e_x^2 e_y \sin(2 (i - \theta_0)) + 13920 e_y^3 \sin(2 (i - \theta_0)) - 
   12540 e_y \sin(4 (i - \theta_0)) \nonumber \\
& - & 13080 e_x^2 e_y \sin(4 (i - \theta_0)) + 
   120 e_y^3 \sin(4 (i - \theta_0)) - 125520 e_x e_y \sin(2 i - \theta_0) \nonumber \\
& - & 
   94260 e_x e_y \sin(4 i - \theta_0) + 451320 e_x e_y \sin(\theta_0) - 60480 e_y \sin(2 \theta_0) \nonumber \\
& - & 
   106080 e_x^2 e_y \sin(2 \theta_0) - 26880 e_y^3 \sin(2 \theta_0) - 
   67920 e_x e_y \sin(3 \theta_0) \nonumber \\
& + & 38760 e_y \sin(4 \theta_0) + 2640 e_x^2 e_y \sin(4 \theta_0) + 
   29040 e_y^3 \sin(4 \theta_0) + 5664 e_x e_y \sin(5 \theta_0) \nonumber \\
& - & 7040 e_y \sin(6 \theta_0) - 
   3040 e_x^2 e_y \sin(6 \theta_0) - 3840 e_y^3 \sin(6 \theta_0) - 5400 e_x e_y \sin(7 \theta_0) \nonumber
\end{eqnarray}
\begin{eqnarray}
& - & 
   990 e_x^2 e_y \sin(8 \theta_0) + 330 e_y^3 \sin(8 \theta_0) - 
   51360 e_y \sin(2 (i + \theta_0)) \nonumber \\
& - & 97440 e_x^2 e_y \sin(2 (i + \theta_0)) - 
   13920 e_y^3 \sin(2 (i + \theta_0)) + 12540 e_y \sin(4 (i + \theta_0)) \nonumber \\
& + & 
   13080 e_x^2 e_y \sin(4 (i + \theta_0)) - 120 e_y^3 \sin(4 (i + \theta_0)) + 
   125520 e_x e_y \sin(2 i + \theta_0) \nonumber \\
& - & 72000 e_y \sin(2 (2 i + \theta_0)) - 
   8880 e_x^2 e_y \sin(2 (2 i + \theta_0)) - 12960 e_y^3 \sin(2 (2 i + \theta_0)) \nonumber \\
& + & 
   94260 e_x e_y \sin(4 i + \theta_0) - 1200 e_y \sin(2 (i + 2 \theta_0)) - 
   2880 e_x^2 e_y \sin(2 (i + 2 \theta_0)) \nonumber \\
& + & 8640 e_y^3 \sin(2 (i + 2 \theta_0)) - 
   405 e_x^2 e_y \sin(4 (i + 2 \theta_0)) + 135 e_y^3 \sin(4 (i + 2 \theta_0)) \nonumber \\
& + & 
   6880 e_y \sin(2 (i + 3 \theta_0)) + 1760 e_x^2 e_y \sin(2 (i + 3 \theta_0)) + 
   2080 e_y^3 \sin(2 (i + 3 \theta_0)) \nonumber \\
& - & 69440 e_x e_y \sin(2 i + 3 \theta_0) + 
   11240 e_x e_y \sin(4 i + 3 \theta_0) + 900 e_x^2 e_y \sin(2 (i + 4 \theta_0)) \nonumber \\
& - & 
   300 e_y^3 \sin(2 (i + 4 \theta_0)) + 288 e_x e_y \sin(2 i + 5 \theta_0) + 
   12240 e_x e_y \sin(4 i + 5 \theta_0) \nonumber \\
& - & 3360 e_y \sin(4 i + 6 \theta_0) + 
   1680 e_x^2 e_y \sin(4 i + 6 \theta_0) - 2080 e_y^3 \sin(4 i + 6 \theta_0) \nonumber \\
& + & 
   5040 e_x e_y \sin(2 i + 7 \theta_0) - 2340 e_x e_y \sin(4 i + 7 \theta_0)),
\end{eqnarray}
for the second order $y$ component of the eccentricity vector $e_{y2}$ is:
\begin{eqnarray}
\overline{e_{y2}} & = & -\frac{1}{81920} A^2 \Bigl(-33600 \sin (2 (i-\theta_0)) e_x^3+6480 \sin (4 (i-\theta_0)) e_x^3\nonumber \\
& + & 27120 \sin (2 \theta_0) e_x^3-1440 \sin (4 \theta_0) e_x^3-3920 \sin (6 \theta_0) e_x^3-330 \sin (8 \theta_0) e_x^3\nonumber \\
& + & 33600 \sin (2 (i+\theta_0)) e_x^3-6480 \sin (4 (i+\theta_0)) e_x^3+840 \sin (2 (2 i+\theta_0)) e_x^3\nonumber \\
& + & 3360 \sin (2 (i+2 \theta_0)) e_x^3-135 \sin (4 (i+2 \theta_0)) e_x^3+1600 \sin (2 (i+3 \theta_0)) e_x^3\nonumber \\
& + & 300 \sin (2 (i+4 \theta_0)) e_x^3-280 \sin (4 i+6 \theta_0) e_x^3-3360 \sin (2 (i-2 \theta_0)) e_x^3\nonumber \\
& + & 135 \sin (4 (i-2 \theta_0)) e_x^3-840 \sin (4 i-2 \theta_0) e_x^3-1600 \sin (2 (i-3 \theta_0)) e_x^3\nonumber \\
& - & 300 \sin (2 (i-4 \theta_0)) e_x^3+280 \sin (4 i-6 \theta_0) e_x^3-55600 e_y e_x^2\nonumber \\
& - & 58080 e_y \cos (2 (i-\theta_0)) e_x^2+13260 e_y \cos (4 (i-\theta_0)) e_x^2\nonumber \\
& - & 74880 e_y \cos (2 \theta_0) e_x^2-12600 e_y \cos (4 \theta_0) e_x^2+4480 e_y \cos (6 \theta_0) e_x^2\nonumber \\
& + & 990 e_y \cos (8 \theta_0) e_x^2-58080 e_y \cos (2 (i+\theta_0)) e_x^2\nonumber \\
& + & 13260 e_y \cos (4 (i+\theta_0)) e_x^2+11040 e_y \cos (2 (2 i+\theta_0)) e_x^2\nonumber \\
& + & 2640 e_y \cos (2 (i+2 \theta_0)) e_x^2+405 e_y \cos (4 (i+2 \theta_0)) e_x^2\nonumber \\
& - & 800 e_y \cos (2 (i+3 \theta_0)) e_x^2-900 e_y \cos (2 (i+4 \theta_0)) e_x^2\nonumber \\
& - & 1440 e_y \cos (4 i+6 \theta_0) e_x^2+2640 e_y \cos (2 (i-2 \theta_0)) e_x^2\nonumber \\
& + & 405 e_y \cos (4 (i-2 \theta_0)) e_x^2+11040 e_y \cos (4 i-2 \theta_0) e_x^2\nonumber \\
& - & 800 e_y \cos (2 (i-3 \theta_0)) e_x^2-900 e_y \cos (2 (i-4 \theta_0)) e_x^2\nonumber \\
& - & 1440 e_y \cos (4 i-6 \theta_0) e_x^2-45960 \sin (2 i-\theta_0) e_x^2-38010 \sin (4 i-\theta_0) e_x^2\nonumber \\
& - & 137220 \sin (\theta_0) e_x^2+11600 \sin (3 \theta_0) e_x^2-21048 \sin (5 \theta_0) e_x^2\nonumber \\
& - & 2700 \sin (7 \theta_0) e_x^2+45960 \sin (2 i+\theta_0) e_x^2+38010 \sin (4 i+\theta_0) e_x^2\nonumber \\
& + & 36560 \sin (2 i+3 \theta_0) e_x^2-8440 \sin (4 i+3 \theta_0) e_x^2+9504 \sin (2 i+5 \theta_0) e_x^2\nonumber \\
& - & 900 \sin (4 i+5 \theta_0) e_x^2+2520 \sin (2 i+7 \theta_0) e_x^2-1170 \sin (4 i+7 \theta_0) e_x^2\nonumber \\
& - & 36560 \sin (2 i-3 \theta_0) e_x^2+8440 \sin (4 i-3 \theta_0) e_x^2-9504 \sin (2 i-5 \theta_0) e_x^2\nonumber \\
& + & 900 \sin (4 i-5 \theta_0) e_x^2-2520 \sin (2 i-7 \theta_0) e_x^2+1170 \sin (4 i-7 \theta_0) e_x^2\nonumber \\
& + & 232080 e_y \cos (2 i-\theta_0) e_x+227460 e_y \cos (4 i-\theta_0) e_x+248280 e_y \cos (\theta_0) e_x\nonumber \\
& - & 43920 e_y \cos (3 \theta_0) e_x+4896 e_y \cos (5 \theta_0) e_x+5400 e_y \cos (7 \theta_0) e_x\nonumber \\
& + & 232080 e_y \cos (2 i+\theta_0) e_x+227460 e_y \cos (4 i+\theta_0) e_x\nonumber
\end{eqnarray}
\begin{eqnarray}
& - & 12160 e_y \cos (2 i+3 \theta_0) e_x+21320 e_y \cos (4 i+3 \theta_0) e_x\nonumber \\
& + & 672 e_y \cos (2 i+5 \theta_0) e_x-10800 e_y \cos (4 i+5 \theta_0) e_x-5040 e_y \cos (2 i+7 \theta_0) e_x\nonumber \\
& + & 2340 e_y \cos (4 i+7 \theta_0) e_x-12160 e_y \cos (2 i-3 \theta_0) e_x\nonumber \\
& + & 21320 e_y \cos (4 i-3 \theta_0) e_x+672 e_y \cos (2 i-5 \theta_0) e_x-10800 e_y \cos (4 i-5 \theta_0) e_x\nonumber \\
& - & 5040 e_y \cos (2 i-7 \theta_0) e_x+2340 e_y \cos (4 i-7 \theta_0) e_x+40320 e_y^2 \sin (2 (i-\theta_0)) e_x\nonumber \\
& - & 99360 \sin (2 (i-\theta_0)) e_x-16080 e_y^2 \sin (4 (i-\theta_0)) e_x+8580 \sin (4 (i-\theta_0)) e_x\nonumber \\
& - & 37680 e_y^2 \sin (2 \theta_0) e_x+77280 \sin (2 \theta_0) e_x+5280 e_y^2 \sin (4 \theta_0) e_x\nonumber \\
& - & 26520 \sin (4 \theta_0) e_x-2800 e_y^2 \sin (6 \theta_0) e_x-7040 \sin (6 \theta_0) e_x+990 e_y^2 \sin (8 \theta_0) e_x\nonumber \\
& - & 40320 e_y^2 \sin (2 (i+\theta_0)) e_x+99360 \sin (2 (i+\theta_0)) e_x+16080 e_y^2 \sin (4 (i+\theta_0)) e_x\nonumber \\
& - & 8580 \sin (4 (i+\theta_0)) e_x+3480 e_y^2 \sin (2 (2 i+\theta_0)) e_x+77040 \sin (2 (2 i+\theta_0)) e_x \nonumber \\
& + & 12000 e_y^2 \sin (2 (i+2 \theta_0)) e_x+29520 \sin (2 (i+2 \theta_0)) e_x\nonumber \\
& + & 405 e_y^2 \sin (4 (i+2 \theta_0)) e_x+3200 e_y^2 \sin (2 (i+3 \theta_0)) e_x+6880 \sin (2 (i+3 \theta_0)) e_x\nonumber \\
& - & 900 e_y^2 \sin (2 (i+4 \theta_0)) e_x-3720 e_y^2 \sin (4 i+6 \theta_0) e_x-3360 \sin (4 i+6 \theta_0) e_x\nonumber \\
& - & 12000 e_y^2 \sin (2 (i-2 \theta_0)) e_x-29520 \sin (2 (i-2 \theta_0)) e_x-405 e_y^2 \sin (4 (i-2 \theta_0)) e_x\nonumber \\
& - & 3480 e_y^2 \sin (4 i-2 \theta_0) e_x-77040 \sin (4 i-2 \theta_0) e_x-3200 e_y^2 \sin (2 (i-3 \theta_0)) e_x\nonumber \\
& - & 6880 \sin (2 (i-3 \theta_0)) e_x+900 e_y^2 \sin (2 (i-4 \theta_0)) e_x+3720 e_y^2 \sin (4 i-6 \theta_0) e_x\nonumber \\
& + & 3360 \sin (4 i-6 \theta_0) e_x+130320 e_y^3+41280 \pi ^2 e_y+24200 e_y\nonumber \\
& + & 160 e_y \left(-539 e_x^2+360 \pi ^2+913 e_y^2-7\right) \cos (2 i)\nonumber \\
& + & 40 e_y \left(-102 e_x^2+600 \pi ^2+3138 e_y^2+2303\right) \cos (4 i)+40800 e_y^3 \cos (2 (i-\theta_0))\nonumber \\
& + & 7200 e_y \cos (2 (i-\theta_0))-9300 e_y^3 \cos (4 (i-\theta_0))-9420 e_y \cos (4 (i-\theta_0))\nonumber \\
& + & 52320 e_y^3 \cos (2 \theta_0)+15360 e_y \cos (2 \theta_0)-19320 e_y^3 \cos (4 \theta_0)-21960 e_y \cos (4 \theta_0)\nonumber \\
& + & 3360 e_y^3 \cos (6 \theta_0)+7040 e_y \cos (6 \theta_0)-330 e_y^3 \cos (8 \theta_0)+40800 e_y^3 \cos (2 (i+\theta_0))\nonumber \\
& + & 7200 e_y \cos (2 (i+\theta_0))-9300 e_y^3 \cos (4 (i+\theta_0))-9420 e_y \cos (4 (i+\theta_0))\nonumber \\
& + & 21360 e_y^3 \cos (2 (2 i+\theta_0))+15840 e_y \cos (2 (2 i+\theta_0))-6000 e_y^3 \cos (2 (i+2 \theta_0))\nonumber \\
& - & 4560 e_y \cos (2 (i+2 \theta_0))-135 e_y^3 \cos (4 (i+2 \theta_0))-2400 e_y^3 \cos (2 (i+3 \theta_0))\nonumber \\
& - & 6880 e_y \cos (2 (i+3 \theta_0))+300 e_y^3 \cos (2 (i+4 \theta_0))+2000 e_y^3 \cos (4 i+6 \theta_0)\nonumber \\
& + & 3360 e_y \cos (4 i+6 \theta_0)-6000 e_y^3 \cos (2 (i-2 \theta_0))-4560 e_y \cos (2 (i-2 \theta_0))\nonumber \\
& - & 135 e_y^3 \cos (4 (i-2 \theta_0))+21360 e_y^3 \cos (4 i-2 \theta_0)+15840 e_y \cos (4 i-2 \theta_0)\nonumber \\
& - & 2400 e_y^3 \cos (2 (i-3 \theta_0))-6880 e_y \cos (2 (i-3 \theta_0))+300 e_y^3 \cos (2 (i-4 \theta_0))\nonumber \\
& + & 2000 e_y^3 \cos (4 i-6 \theta_0)+3360 e_y \cos (4 i-6 \theta_0)-234840 e_y^2 \sin (2 i-\theta_0)\nonumber \\
& - & 46560 \sin (2 i-\theta_0)-102750 e_y^2 \sin (4 i-\theta_0)-65160 \sin (4 i-\theta_0)\nonumber \\
& + & 353940 e_y^2 \sin (\theta_0)+99120 \sin (\theta_0)+63920 e_y^2 \sin (3 \theta_0)+1280 \sin (3 \theta_0)\nonumber \\
& - & 16152 e_y^2 \sin (5 \theta_0)-6000 \sin (5 \theta_0)+2700 e_y^2 \sin (7 \theta_0)+234840 e_y^2 \sin (2 i+\theta_0)\nonumber \\
& + & 46560 \sin (2 i+\theta_0)+102750 e_y^2 \sin (4 i+\theta_0)+65160 \sin (4 i+\theta_0)\nonumber \\
& + & 18160 e_y^2 \sin (2 i+3 \theta_0)+32320 \sin (2 i+3 \theta_0)+40120 e_y^2 \sin (4 i+3 \theta_0)\nonumber \\
& - & 9920 \sin (4 i+3 \theta_0)+10176 e_y^2 \sin (2 i+5 \theta_0)+6240 \sin (2 i+5 \theta_0)\nonumber \\
& - & 11700 e_y^2 \sin (4 i+5 \theta_0)-3240 \sin (4 i+5 \theta_0)-2520 e_y^2 \sin (2 i+7 \theta_0)\nonumber
\end{eqnarray}
\begin{eqnarray}
& + & 1170 e_y^2 \sin (4 i+7 \theta_0)-18160 e_y^2 \sin (2 i-3 \theta_0)-32320 \sin (2 i-3 \theta_0)\nonumber \\
& - & 40120 e_y^2 \sin (4 i-3 \theta_0)+9920 \sin (4 i-3 \theta_0)-10176 e_y^2 \sin (2 i-5 \theta_0)\nonumber \\
& - & 6240 \sin (2 i-5 \theta_0)+11700 e_y^2 \sin (4 i-5 \theta_0)+3240 \sin (4 i-5 \theta_0)\nonumber \\
& + & 2520 e_y^2 \sin (2 i-7 \theta_0)-1170 e_y^2 \sin (4 i-7 \theta_0)\Bigl),
\end{eqnarray}
for the second order solution of the inclination $i_2$ is:
\begin{eqnarray}
\overline{i_2} & = & \frac{1}{1024} A^2 \sin (2 i) \Bigl(\left(e_y^2-e_x^2\right) \cos (2 (i-3 \theta_0))+249 e_x^2 \cos (2 (i-\theta_0))\nonumber \\
& + & 249 e_x^2 \cos (2 (i+\theta_0))+36 e_x^2 \cos (2 (i+2 \theta_0))-e_x^2 \cos (2 (i+3 \theta_0))\nonumber \\
& + & 36 e_x^2 \cos (2 (i-2 \theta_0))-102 e_x e_y \sin (2 (i-\theta_0))+102 e_x e_y \sin (2 (i+\theta_0))\nonumber \\
& + & 42 e_x e_y \sin (2 (i+2 \theta_0))-2 e_x e_y \sin (2 (i+3 \theta_0))-42 e_x e_y \sin (2 (i-2 \theta_0))\nonumber \\
& + & 2 e_x e_y \sin (2 (i-3 \theta_0))+552 e_x \cos (2 i-\theta_0)+552 e_x \cos (2 i+\theta_0)\nonumber \\
& + & 216 e_x \cos (2 i+3 \theta_0)+216 e_x \cos (2 i-3 \theta_0)+15 e_y^2 \cos (2 (i-\theta_0))\nonumber \\
& + & 15 e_y^2 \cos (2 (i+\theta_0))-6 e_y^2 \cos (2 (i+2 \theta_0))+e_y^2 \cos (2 (i+3 \theta_0))\nonumber \\
& - & 6 e_y^2 \cos (2 (i-2 \theta_0))-120 e_y \sin (2 i-\theta_0)+120 e_y \sin (2 i+\theta_0)\nonumber \\
& + & 120 e_y \sin (2 i+3 \theta_0)-120 e_y \sin (2 i-3 \theta_0)+216 \cos (2 (i-\theta_0))\nonumber \\
& + & 216 \cos (2 (i+\theta_0))+6 \cos (2 (i+2 \theta_0))+6 \cos (2 (i-2 \theta_0))\nonumber \\
& + & 16 \cos (2 i) \left(15 e_x^2+45 e_y^2+19\right)+54 e_x^2 \cos (2 \theta_0)-48 e_x^2 \cos (4 \theta_0)\nonumber \\
& - & 6 e_x^2 \cos (6 \theta_0)+228 e_x e_y \sin (2 \theta_0)+60 e_x e_y \sin (4 \theta_0)-12 e_x e_y \sin (6 \theta_0)\nonumber \\
& + & 432 e_x \cos (\theta_0)-48 e_x \cos (5 \theta_0)+378 e_y^2 \cos (2 \theta_0)-108 e_y^2 \cos (4 \theta_0)\nonumber \\
& + & 6 e_y^2 \cos (6 \theta_0)-624 e_y \sin (\theta_0)+480 e_y \sin (3 \theta_0)-48 e_y \sin (5 \theta_0)\nonumber \\
& + & 336 \cos (2 \theta_0)-84 \cos (4 \theta_0)-368 e_x^2-16 e_y^2-160\Bigl),
\end{eqnarray}
and for the second order of the right ascension of the ascending node $\Omega_2$ is:
\begin{eqnarray}
\overline{\Omega_2} & = & 
\frac{1}{512} A^2 \cos (i) \Bigl(-9 e_x^2 \sin (2 (i-\theta_0))
+9 e_x^2 \sin (2 (i+\theta_0))\nonumber \\
& - & 75 e_x^2 \sin (2 (i+2 \theta_0))
-3 e_x^2 \sin (2 (i+3 \theta_0))
+75 e_x^2 \sin (2 (i-2 \theta_0))\nonumber \\
& + & 3 e_x^2 \sin (2 (i-3 \theta_0))
+90 e_x e_y \cos (2 (i-\theta_0))
+90 e_x e_y \cos (2 (i+\theta_0))\nonumber \\
& + & 156 e_x e_y \cos (2 (i+2 \theta_0))
+6 e_x e_y \cos (2 (i+3 \theta_0))\nonumber \\
& + & 156 e_x e_y \cos (2 (i-2 \theta_0))
+6 e_x e_y \cos (2 (i-3 \theta_0))
-144 e_x \sin (2 i-\theta_0)\nonumber \\
& + & 144 e_x \sin (2 i+\theta_0)
-28 e_x \sin (2 i+3 \theta_0)
-12 e_x \sin (2 i+5 \theta_0)\nonumber \\
& + & 28 e_x \sin (2 i-3 \theta_0)
+12 e_x \sin (2 i-5 \theta_0)
-39 e_y^2 \sin (2 (i-\theta_0))\nonumber \\
& + & 39 e_y^2 \sin (2 (i+\theta_0))
+81 e_y^2 \sin (2 (i+2 \theta_0))
+3 e_y^2 \sin (2 (i+3 \theta_0))\nonumber \\
& - & 81 e_y^2 \sin (2 (i-2 \theta_0))
-3 e_y^2 \sin (2 (i-3 \theta_0))
+2304 e_y \cos (2 i-\theta_0)\nonumber \\
& + & 2304 e_y \cos (2 i+\theta_0)
+52 e_y \cos (2 i+3 \theta_0)
+12 e_y \cos (2 i+5 \theta_0)\nonumber \\
& + & 52 e_y \cos (2 i-3 \theta_0)
+12 e_y \cos (2 i-5 \theta_0)
-504 \sin (2 (i-\theta_0))\nonumber \\
& + & 504 \sin (2 (i+\theta_0))
-12 \sin (2 (i+2 \theta_0))
+12 \sin (2 (i-2 \theta_0))\nonumber \\
& + & 96 e_x e_y \cos (2 i)
+414 e_x^2 \sin (2 \theta_0)
+126 e_x^2 \sin (4 \theta_0)
-10 e_x^2 \sin (6 \theta_0)\nonumber
\end{eqnarray}
\begin{eqnarray}
& - & 660 e_x e_y \cos (2 \theta_0)
-456 e_x e_y \cos (4 \theta_0)
+20 e_x e_y \cos (6 \theta_0)\nonumber \\
& - & 1248 e_x \sin (\theta_0)
+536 e_x \sin (3 \theta_0)
-72 e_x \sin (5 \theta_0)
+450 e_y^2 \sin (2 \theta_0)\nonumber \\
& - & 330 e_y^2 \sin (4 \theta_0)
+10 e_y^2 \sin (6 \theta_0)
-576 e_y \cos (\theta_0)
-1160 e_y \cos (3 \theta_0)\nonumber \\
& + & 72 e_y \cos (5 \theta_0)
+336 \sin (2 \theta_0)
-120 \sin (4 \theta_0)
-1504 e_x e_y\Bigl).
\end{eqnarray}


\section{Examples of application}
\label{sec:examples}

In order to assess the accuracy of both the approximate solution and the proposed transformation, in this section four examples of application are covered, the first one for a near circular orbit, the second one for a high eccentric orbit, the third one for a hyperbolic orbit, and the last one for a parabolic orbit. In all cases, an orbit around the Earth is considered.

\subsection{Near circular orbit}

\begin{figure}[h!]
	\centering
	{\includegraphics[width = \textwidth]{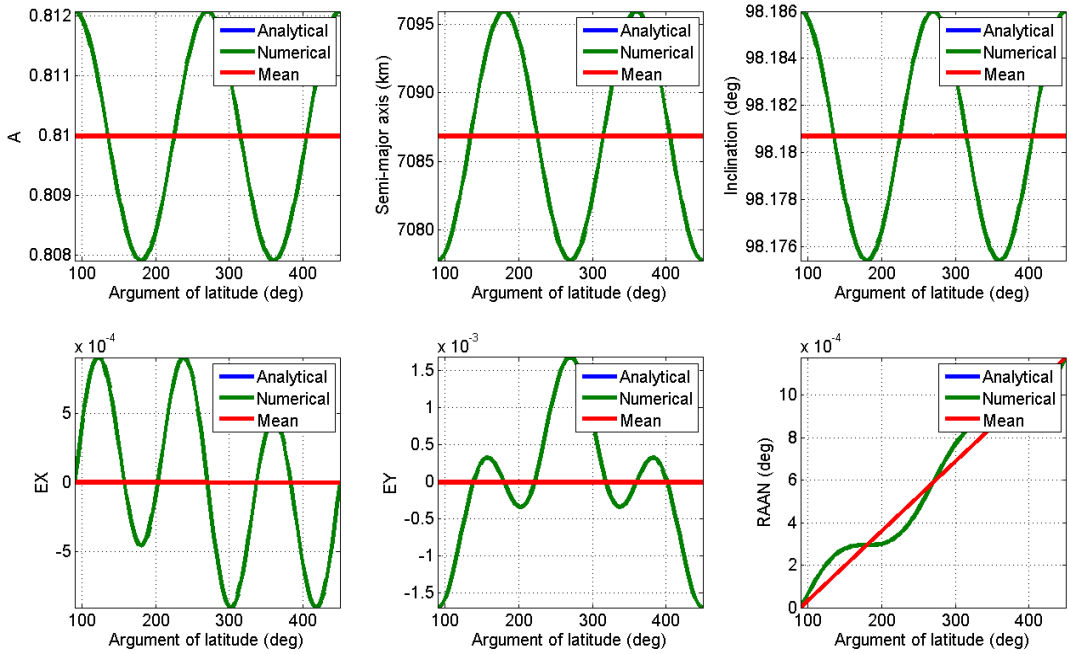}}
	\caption{Osculating and mean orbital elements for the frozen sun-synchronous orbit.}
	\label{fig:mean_circ}
\end{figure}

As an example of near circular orbit, a sun-synchronous frozen orbit has been selected with a mean semi-major axis of $7086.864$ km. To that end, the initial orbital elements chosen are the following: $A_0 = 0.812$, $e_{x0} = 0.0$, $e_{y0} = -0.001696$, $i_0 = 98.186$ deg, $\Omega_0 = 0.0$ deg, and $\theta_0 = 90$ deg. Figure~\ref{fig:mean_circ} shows the orbital element evolution for this set of initial conditions for a complete orbital revolution. This includes both the numerical and analytical approximate solutions, as well as the mean values of each orbital element at every instant. In this regard, the computation of the evolution in the mean value of the orbital elements has been done by performing the transformation from osculating to mean elements provided in this work applied to the numerical solution. As can be seen, the transformations work as intended and clearly show the frozen character of the orbit selected. Note that this transformation can be also performed from the approximate analytical solution obtaining a very similar result.  

\begin{figure}[h!]
	\centering
	{\includegraphics[width = \textwidth]{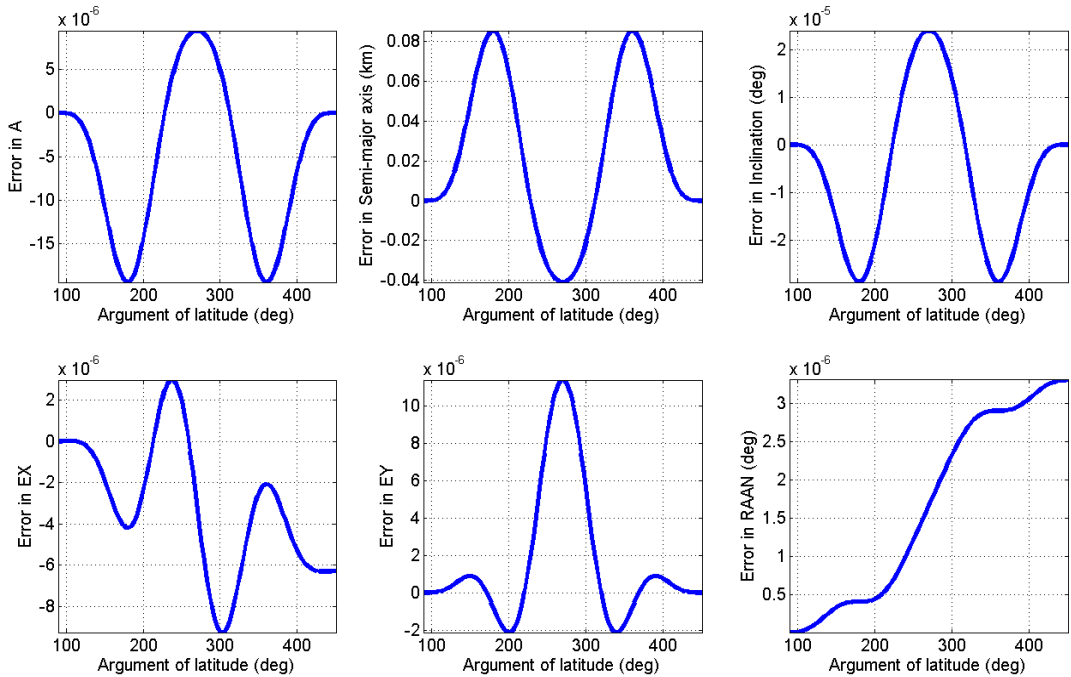}}
	\caption{Error of the first order solution for the frozen sun-synchronous orbit.}
	\label{fig:error_circ_o1}
\end{figure}

More important than the evolution of the orbital variables is probably the assessment of the accuracy of the approximate analytical solution. To that end, Figure~\ref{fig:error_circ_o1} shows the error of the first order approximation presented in this manuscript. As can be seen, even for a first order solution, a good accuracy is obtained, having maximum errors in position for one orbit with an order of magnitude of 100 meters. Note that the error of the solution gets close to zero after the object has completed a full revolution. This helps maintaining the accuracy in longer propagation times.

\begin{figure}[h!]
	\centering
	{\includegraphics[width = 0.55\textwidth]{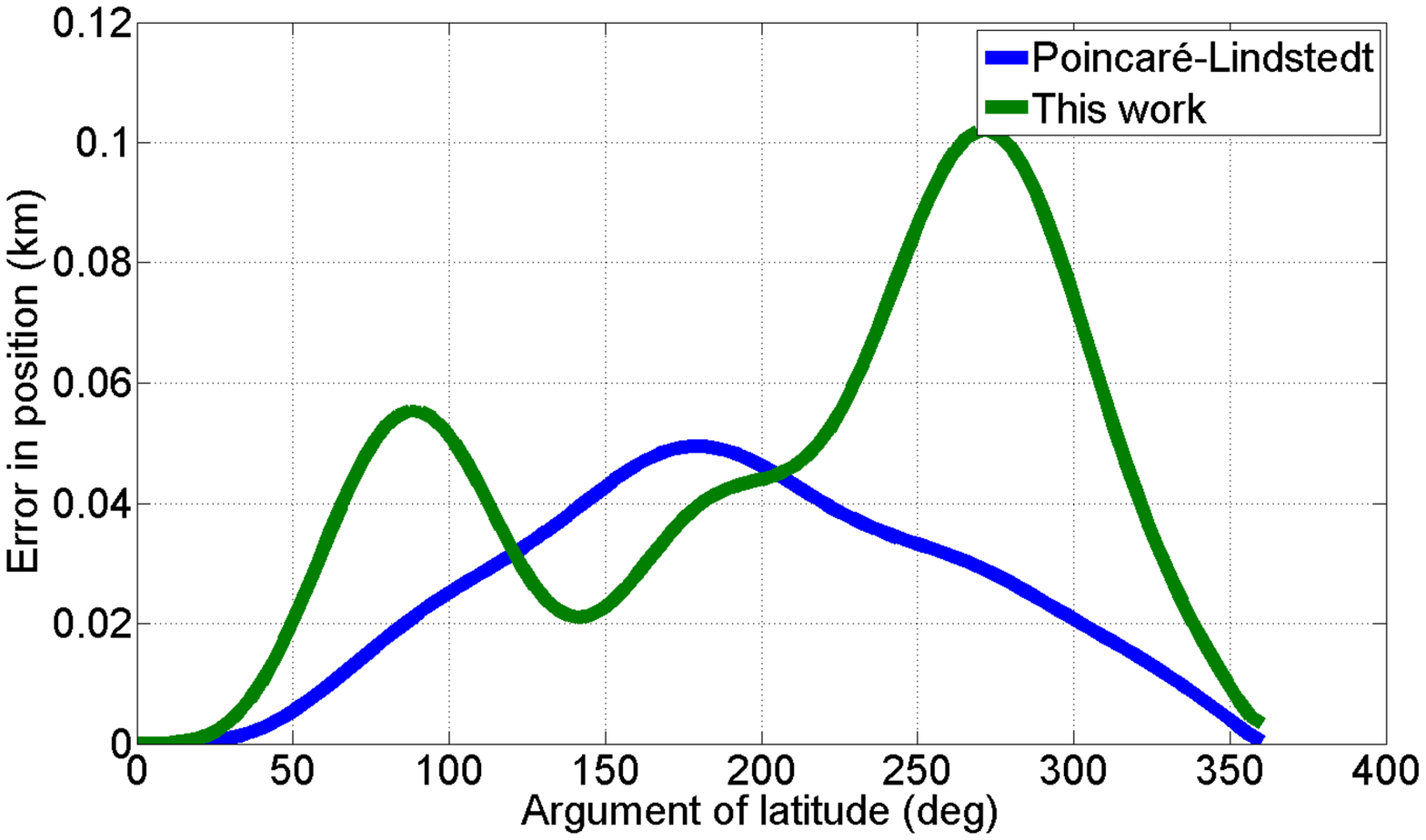}}
	\caption{First order error comparison between the Poincar\'e-Lindstedt method and the perturbation method proposed in this work.}
	\label{fig:comparison_circ}
\end{figure}

Additionally, it is worth including the comparison of this first order analytical solution with other perturbation methods. To that end, the Poincar\'e-Lindstedt method proposed in Arnas and Linares~\cite{zonal} has been selected, as it also directly provides an osculating variation of the orbital elements. The results of that comparison in position error can be seen in Figure~\ref{fig:comparison_circ}. From them, we can notice that both analytical solutions have the same order of magnitude in error, being the solution provided by the Poincar\'e-Lindstedt method slightly more accurate. Similar behaviour has been found in orbits with higher eccentricity. This is a expected result as the Poincar\'e-Lindstedt method provides a better approximation due to its control in the frequencies of the solution. Nevertheless, the advantages of the proposed method is that it is easier to generate higher order solutions, which overcomes this difference in these short term propagations.

\begin{figure}[h!]
	\centering
	{\includegraphics[width = \textwidth]{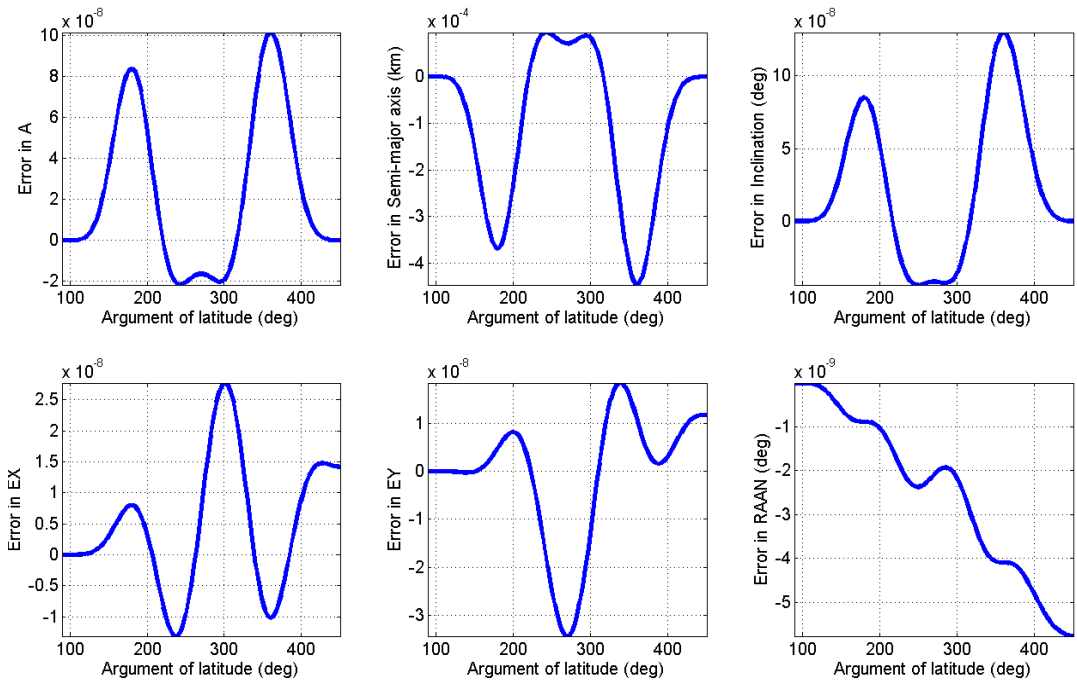}}
	\caption{Error of the second order solution for the frozen sun-synchronous orbit.}
	\label{fig:error_circ_o2}
\end{figure}

As an example of that, Fig.~\ref{fig:error_circ_o2} shows the error associated with the second order approximate solution proposed in this work. As can be seen, and compared with the previous first order solutions, the magnitude of the error has been reduced to nearly three orders of magnitude, having a maximum error in position of 50 cm for one orbital revolution. 

\begin{figure}[h!]
	\centering
	{\includegraphics[width = \textwidth]{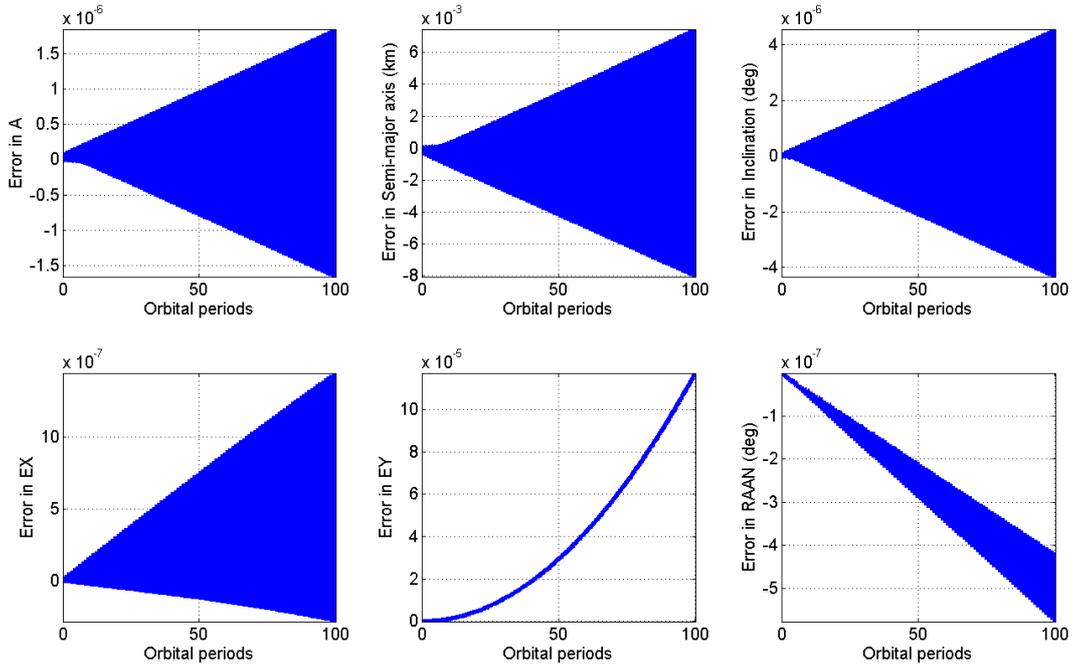}}
	\caption{Long-term error of the second order solution for the frozen sun-synchronous orbit.}
	\label{fig:error_circ_o2_long}
\end{figure}

Finally, Fig.~\ref{fig:error_circ_o2_long} shows the error evolution of the second order solution after 100 orbital revolutions. As can be seen, the error has increased one order of magnitude with respect to the solution during the first orbital revolution. Nevertheless, even with this increase in error, the solution is still very accurate. In this regard, it is important to note that this is an expected result not only of this perturbation method, but also in any other analytical perturbation technique. Additionally, it is worth remember that when focusing on the transformation between osculating and mean elements, this increase in error does not affect that variable transformation since the averaging of the solution is done during a single orbital revolution. 

\subsection{High eccentric orbit}

\begin{figure}[h!]
	\centering
	{\includegraphics[width = \textwidth]{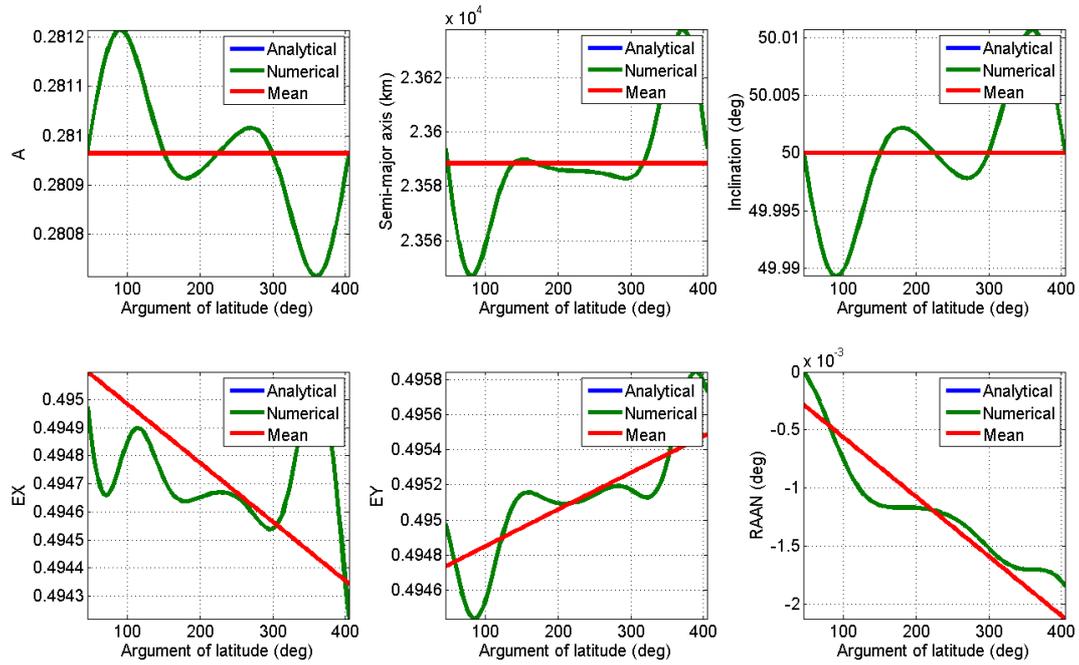}}
	\caption{Osculating and mean orbital elements for the high eccentric orbit.}
	\label{fig:mean_ecc}
\end{figure}

As an example of high eccentric orbit, an orbit presenting apsidal precession has been selected in order to show how well the solution behaves in a more complex dynamic. This corresponds to a worse case scenario of the perturbation method proposed due to the more important non-linear variation of the orbital elements, specially the components of the eccentricity vector. To show this clearly, we select an orbit with an eccentricity of 0.7, inclination of 50 deg, argument of perigee of $45$ deg, and that presents and altitude of 700 km when passing through the perigee of the orbit. This leads to the following initial set of orbital elements: $A_0 = 0.3354$, $e_{x0} = 0.49497$, $e_{y0} = 0.49497$, $i_0 = 50.0$ deg, $\Omega_0 = 0.0$ deg, and $\theta_0 = 45$ deg. Figure~\ref{fig:mean_ecc} shows the evolution of the analytical and numerical solutions, and also includes the result from the transformation from osculating to mean elements. As can be seen, the analytical approximations follow accurately the evolution of the orbital elements under $J_2$ perturbation.

\begin{figure}[h!]
	\centering
	{\includegraphics[width = \textwidth]{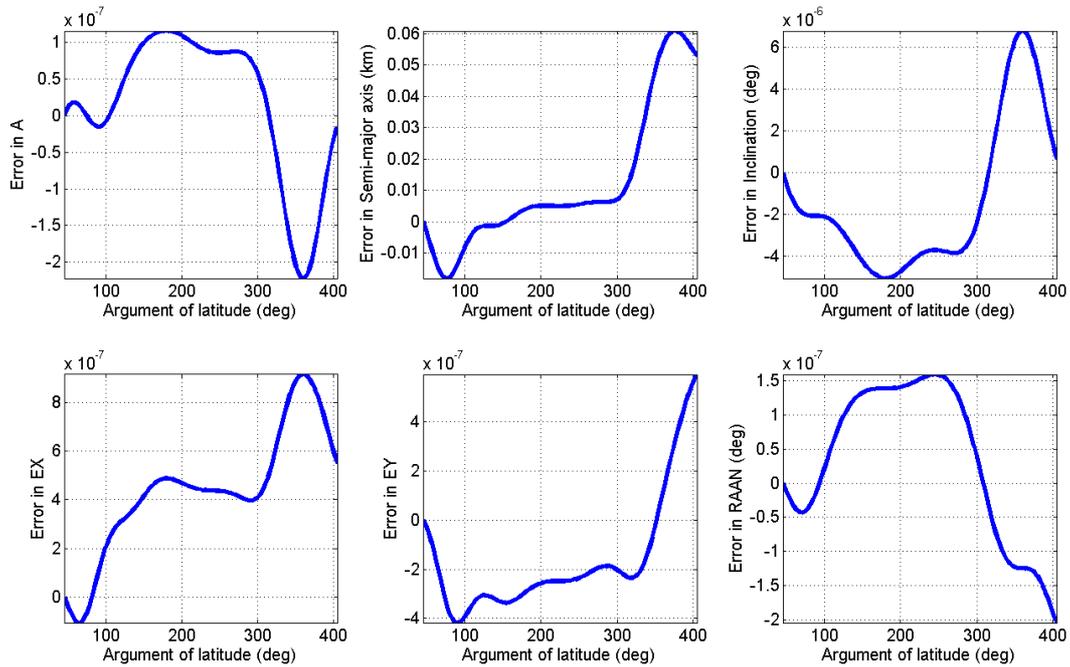}}
	\caption{Error of the first order solution for the high eccentric orbit.}
	\label{fig:error_ecc_o1}
\end{figure}

On the other hand, Fig.~\ref{fig:error_ecc_o1} shows the error of the first order analytical solution. This corresponds to a maximum error of 22 meters that happens close to the apogee of the orbit. This shows the accuracy of the analytical approach even at a first order solution.

\begin{figure}[h!]
	\centering
	{\includegraphics[width = \textwidth]{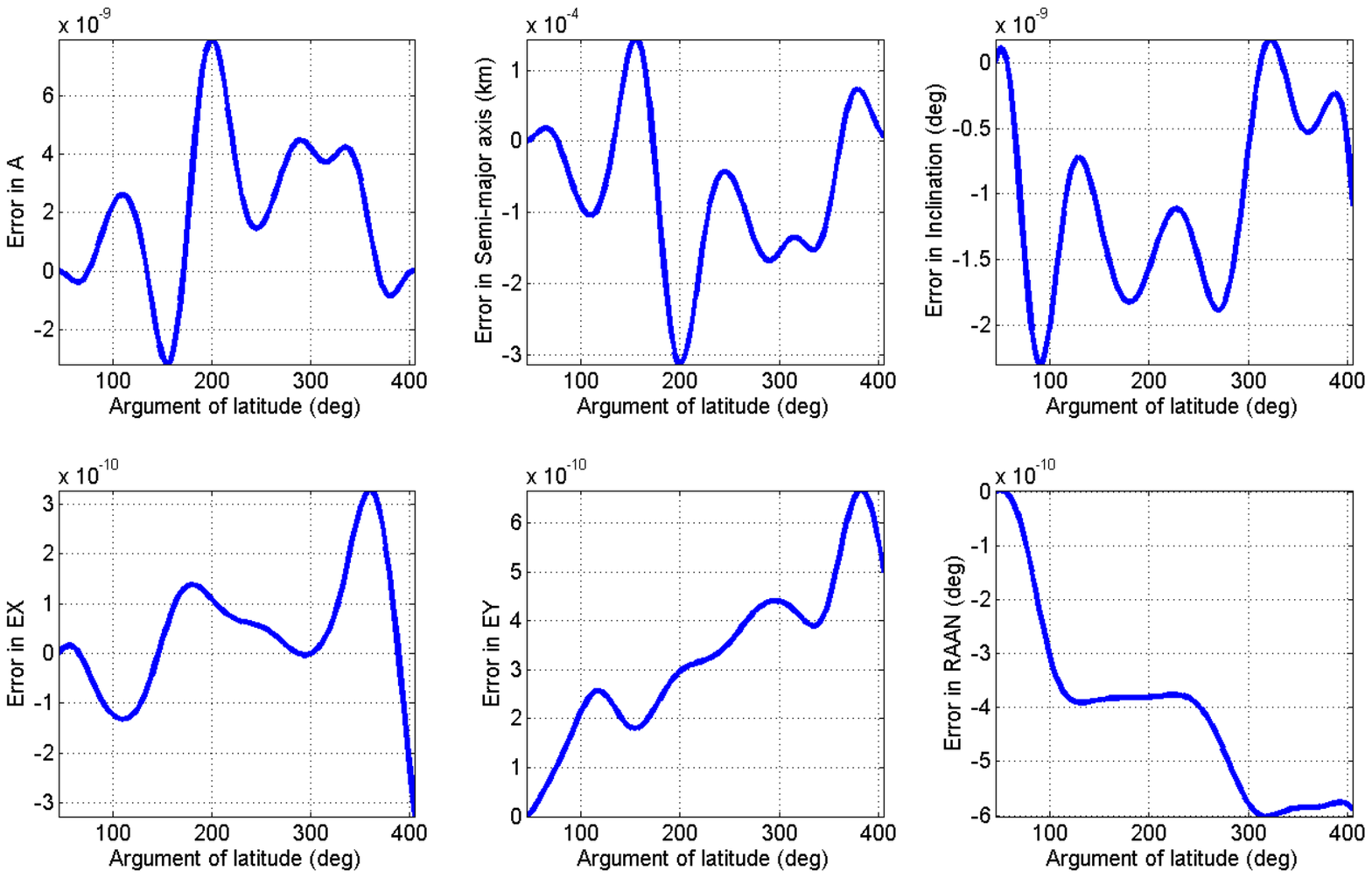}}
	\caption{Error of the second order solution for the high eccentric orbit.}
	\label{fig:error_ecc_o2}
\end{figure}

Additionally, and as done before, a second order solution can be applied. Figure~\ref{fig:error_ecc_o2} shows the resultant error of this approximate solution when compared with the numerical solution. As can be seen, the error has significantly improved with respect to the first order solution, presenting a maximum error of $40$ cm in this first orbital revolution. 

\newpage

\begin{figure}[h!]
	\centering
	{\includegraphics[width = \textwidth]{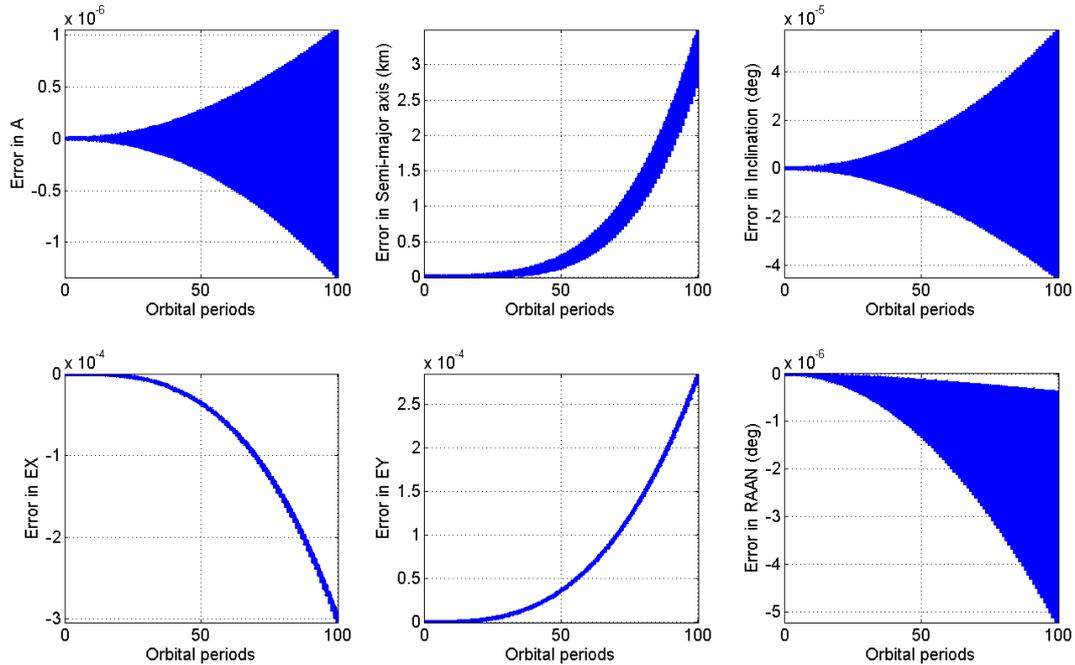}}
	\caption{Long-term error of the second order solution for the high eccentric orbit.}
	\label{fig:error_ecc_o2_long}
\end{figure}

\begin{figure}[h!]
	\centering
	{\includegraphics[width = \textwidth]{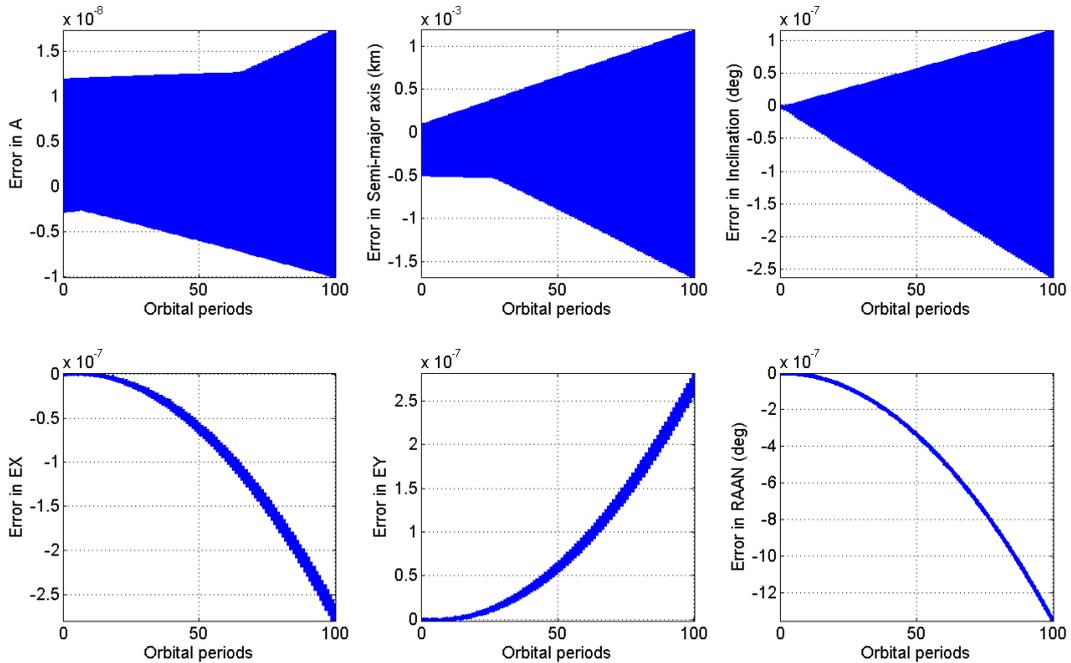}}
	\caption{Long-term error of the second order solution for the high eccentric orbit close to the frozen condition.}
	\label{fig:error_ecc_o2_long2}
\end{figure}

This solution can be extended to a longer term propagation. To that end, Fig.~\ref{fig:error_ecc_o2_long} shows the error of this propagation. In here, we can observe a drastic increase in error when compared with the near circular orbit from Fig.~\ref{fig:error_circ_o2_long}. This effect is due to the lack of control in the frequency of the analytical solution, which becomes more different to the unperturbed frequency the farther the orbit is from one of the frozen conditions. To show this effect clearer, Fig.~\ref{fig:error_ecc_o2_long2} shows the second order error of an orbit with the same orbital elements but its initial inclination, which has been chosen to be $63.43$ deg (very close to the critical inclination). As can be seen, the error has decreased quite significantly, having an error of less than 20 meters in position after the 100 period propagation. This improves in two orders of magnitude the previous orbit. This means that, in general, the accuracy of the methodology gets degraded as orbits get further away form the frozen orbits (either the ones with low eccentricity, or the ones close to the critical inclination), meaning that higher order solutions are required if we want to maintain the original accuracy since they are able to better represent the non-linearities of the solution. This limitation is not important for short-term propagation (several orbital revolutions) as it has been seen, but for longer-term propagations, this effect can be important depending on the specific orbit selected. Nevertheless, this analytical approach behaves specially well for close to frozen orbits (usually the ones that interest us the most for engineering applications), and for the specific case of the transformation between osculating to mean elements since it relies on a propagation for one orbital period.

\subsection{Hyperbolic orbit}

\begin{figure}[h!]
	\centering
	{\includegraphics[width = \textwidth]{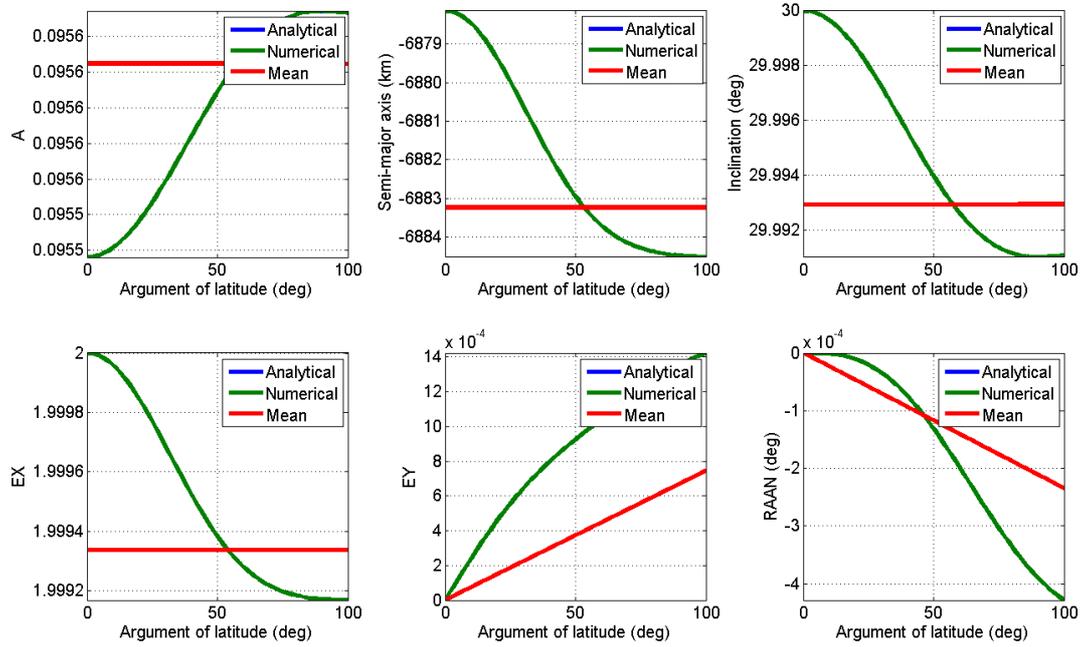}}
	\caption{Osculating and mean orbital elements for the hyperbolic orbit.}
	\label{fig:mean_hyp}
\end{figure}

For the example of hyperbolic orbit, the following initial orbital elements have been selected: $A_0 = 0.092$, $e_{x0} = 2.0$, $e_{y0} = 0.0$, $i_0 = 30.0$ deg, $\Omega_0 = 0.0$ deg, and $\theta_0 = 0.0$ deg. This corresponds to an orbit with perigee at 500 km of altitude over the Earth equator. Figure~\ref{fig:mean_hyp} shows the orbital element evolution for this set of initial orbital elements, as well as the computed value of the mean elements using the proposed methodology and the results from the numerical solution. In this case, the propagation has been stopped at $\theta = 100$ deg, where the satellite was at more than 26000 km of altitude over Earth's surface. As can be seen, even for hyperbolic orbits, the analytical solution is able to accurate represent the osculating and mean behavior of the system.

\begin{figure}[h!]
	\centering
	{\includegraphics[width = \textwidth]{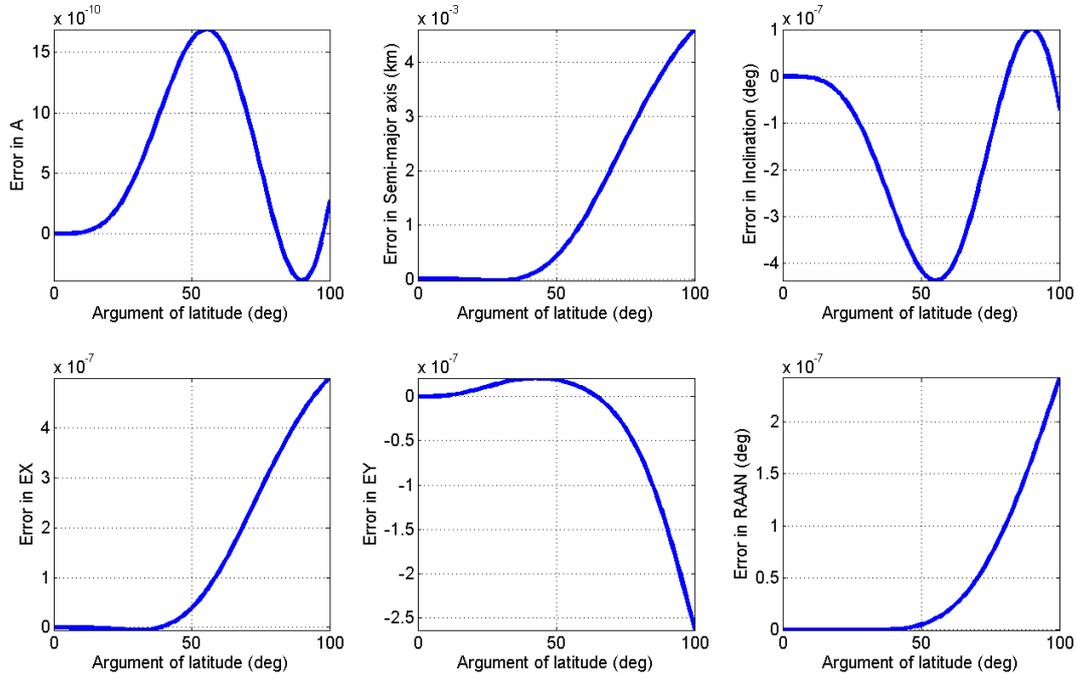}}
	\caption{Error of the first order solution for the hyperbolic orbit.}
	\label{fig:error_hyp_o1}
\end{figure}

\begin{figure}[h!]
	\centering
	{\includegraphics[width = \textwidth]{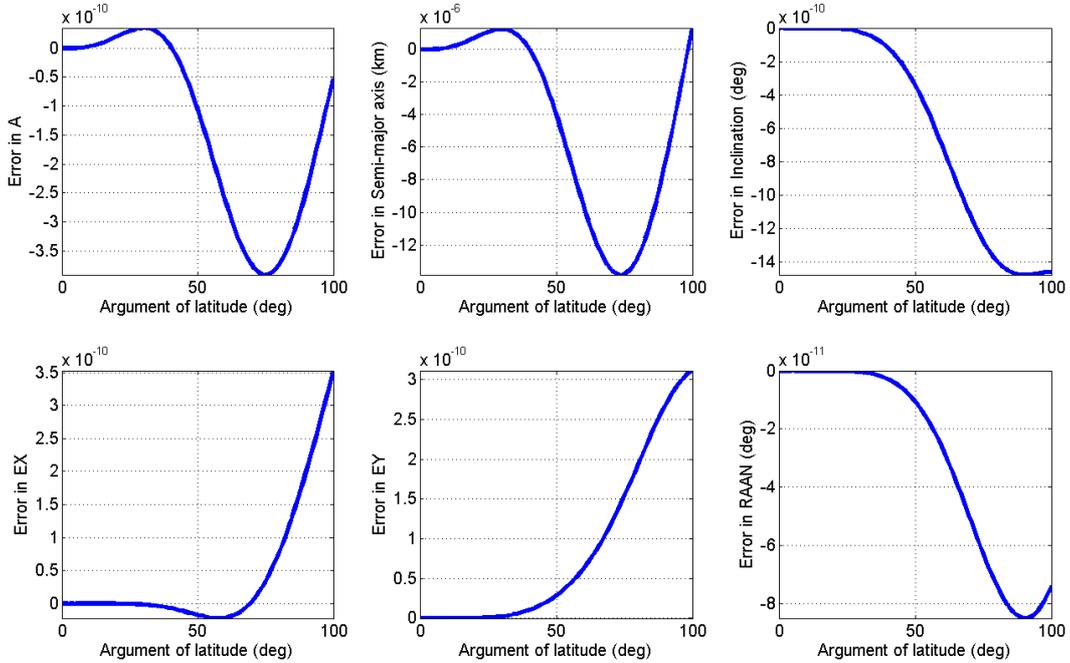}}
	\caption{Error of the second order solution for the hyperbolic orbit.}
	\label{fig:error_hyp_o2}
\end{figure}

Finally, Figs.~\ref{fig:error_hyp_o1} and~\ref{fig:error_hyp_o2}  show the first and second order errors associated with the analytical approximation proposed in this work. As can be seen, the error is very small compared with the large variation that the orbital elements experience. In particular, for the second order solution, a maximum error in position of 60 cm was found, corresponding to the farthest point in the orbit that has been considered. This shows that the proposed methodology can also be effectively applied to hyperbolic orbits in order to obtain either their propagation, or their transformation from osculating to mean elements.

\subsection{Parabolic orbit}

\begin{figure}[h!]
	\centering
	{\includegraphics[width = \textwidth]{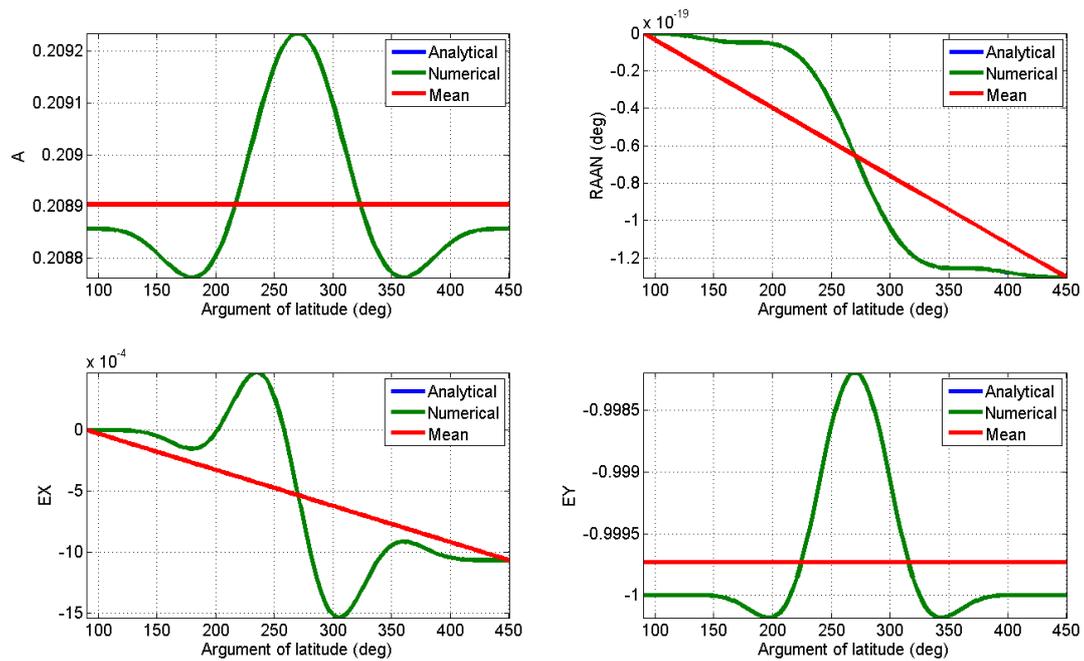}}
	\caption{Osculating and mean orbital elements for the parabolic orbit.}
	\label{fig:mean_par}
\end{figure}

For this final example, a parabolic orbit is selected. In that regard, and in order to be sure that a real parabolic orbit is defined, we set the initial conditions in the farthest point of the orbit, that is, on the infinite, such that the specific energy of the orbit is zero. To better test the effect of $J_2$, we impose that the unperturbed parabolic orbit has a periapsis at 600 km of altitude. Additionally, an inclination of 90 deg is selected with an argument of perigee of 270 deg. This leads to the following initial conditions: $A_0 = 0.2089$, $e_{x0} = 0.0$, $e_{y0} = -1.0$, $i_0 = 90.0$ deg, $\Omega_0 = 0.0$ deg, and $\theta_0 = 90$ deg. Figure~\ref{fig:mean_par} shows the orbital elements evolution and mean values for this orbit. Note that under these initial conditions, the osculating inclination of the orbit remains constant and the semi-major axis is not well defined. For these reasons, these two orbital elements are not represented in this figure. Additionally, even if describing a parabolic orbit is unrealistic, the results for the complete orbit have been included to better show the performance of the proposed methodology. As can be seen in the figure, the analytical solution is able to represent both the osculating behaviour of the system and its mean secular variation. Another interesting point to mention is that the osculating orbit transitions from parabolic to hyperbolic, and later from hyperbolic to elliptic as the orbiting object gets closer to the main celestial body, having the minimum eccentricity in the periapsis of the orbit.

\begin{figure}[h!]
	\centering
	{\includegraphics[width = \textwidth]{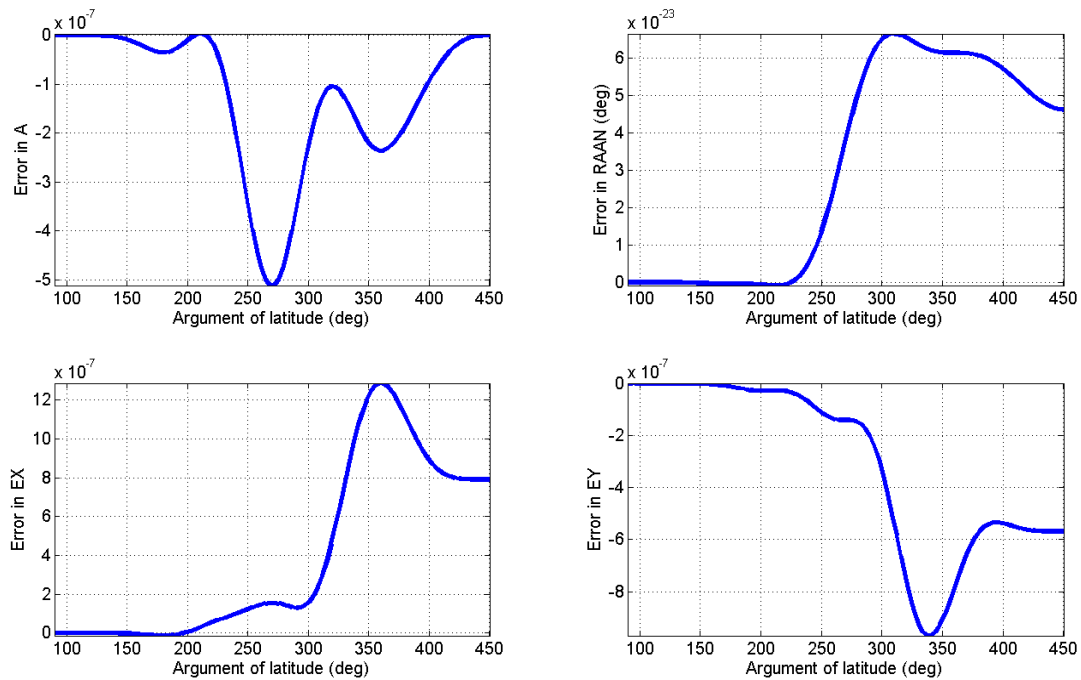}}
	\caption{Error of the first order solution for the parabolic orbit.}
	\label{fig:error_par_o1}
\end{figure}

\begin{figure}[h!]
	\centering
	{\includegraphics[width = \textwidth]{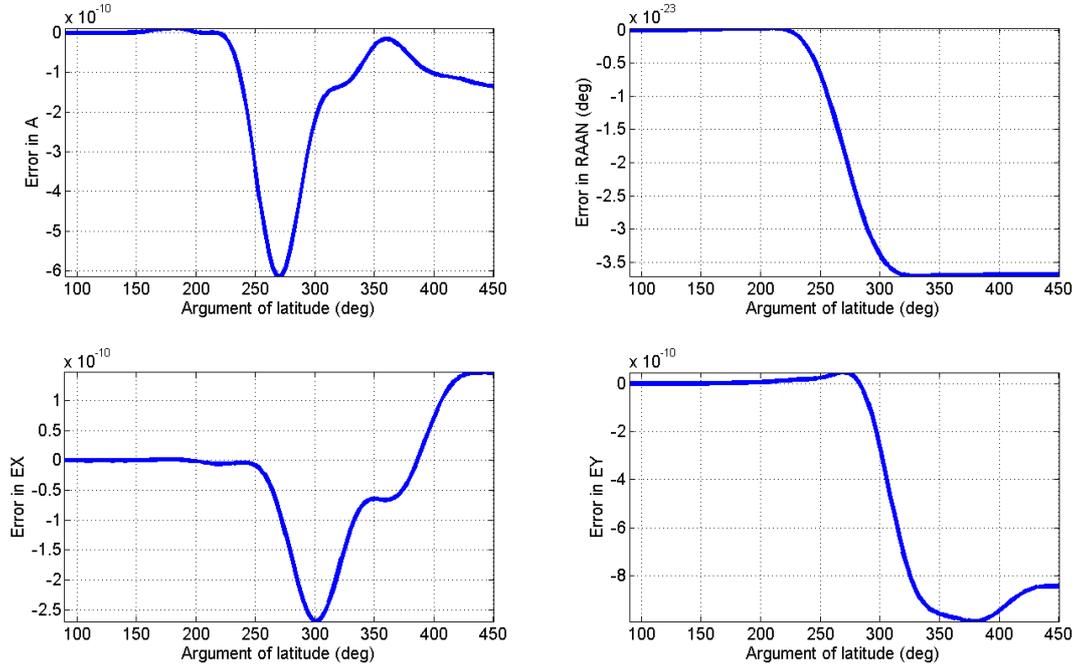}}
	\caption{Error of the second order solution for the parabolic orbit.}
	\label{fig:error_par_o2}
\end{figure}

Additionally, Figs.~\ref{fig:error_par_o1} and~\ref{fig:error_par_o2} present the error of the first and second order solutions when compared with the numerical propagation. As can be seen, the error is very small and with an order of magnitude similar to the one obtained both in hyperbolic and high eccentric orbits. This exemplifies that the proposed methodology can effectively be used to study parabolic orbits.

\section{Conclusions}

This work proposes a perturbation method to study the effects of the oblatness of a celestial body in the dynamics of an orbiting object. This perturbation method is based on performing a time regularization using the geometrical argument of latitude of the orbit, and expanding in a power series of the small parameter both all the orbital elements and the differential equation itself in order to obtain approximate solutions to the problem. This is done with the final goal of obtaining an analytical transformation from osculating to mean elements for orbits at any eccentricity value, including near circular, elliptic, parabolic and hyperbolic orbits.

The proposed methodology has shown good performance for orbits at any eccentricity. This is specially true for short-term propagations, that is, for orbital propagations taking place in several orbital revolutions. For long-term propagations, the approach still remains accurate, but starts to degrade its error performance as orbits get farther away from the frozen orbits. One of the reasons for this effect is the lack of control on the perturbed frequency of the solution. This has been checked through a comparison with the Poincar\'e-Lindstedt method applied to the same problem. Nevertheless, this loss of accuracy compared with the Poincar\'e-Lindstedt method in long-term propagations is balanced by the fact that this approach allows to obtain more reasonably-sized expressions for the second order solution. Additionally, this also means that the proposed approach is specially useful to study frozen orbits due to the mentioned behaviour.

The approximate solution obtained from this perturbation approach is then used to obtain an analytical transformation from osculating to mean elements for orbits at any eccentricity. The results show that this transformation accurately represents the mean behaviour of the system independently of the orbit state selected. Moreover, since the transformation is based on a propagation in a single orbital revolution, the proposed approach is specially suited for this application. This transformation can be useful for many applications, including orbit determination, space situation awareness, mission design, or in combination with other perturbation methods that rely on mean elements instead of osculating elements.

Finally, this work also includes how to define and study parabolic orbits under $J_2$ perturbation, a problem that is difficult to assess using perturbation methods based on averaged methodologies. This shows another potential application of this simple perturbation approach.


\section*{Acknowledgments}

The author wants to thank his parents Silvestre Jos\'e Arnas Escart\'in and Mar\'ia Pilar Mart\'inez Ballesteros for their support while developing this work. The research contained in this document would not have been possible without their continuous encouragement. Gracias por todo.


\end{document}